\documentclass[journal,10pt,twocolumn]{IEEEtran}
\usepackage{caption}
\usepackage{amssymb, amsmath, amsthm}
\usepackage{amsfonts, bbm}
\usepackage{graphicx}
\usepackage{epstopdf}
\usepackage{times}
\usepackage{float} 
\usepackage{array}
\usepackage{arydshln}
\usepackage{bibentry}
\usepackage{booktabs}

\usepackage{cite}
\usepackage{subfigure}
\usepackage{enumitem}
\usepackage{color}
\usepackage{setspace}
\usepackage{bigstrut}
\usepackage{multirow}

\allowdisplaybreaks
\theoremstyle{plain}

\theoremstyle{plain}
\newtheorem{corollary}{Corollary}
\theoremstyle{plain}
\newtheorem{lemmacounter}{Theorem}
\newtheorem{lemma}[lemmacounter]{Lemma}
\theoremstyle{plain}

\theoremstyle{plain}

\newtheorem{special case}[defcounter]{Special Case}
 
\newcommand*{\Scale}[2][4]{\scalebox{#1}{$#2$}}%

\newcommand{\figref}[1]{Fig.~\ref{#1}}
\begin{document}

\title{Analysis of Massive MIMO-Enabled Downlink Wireless Backhauling for Full-Duplex Small Cells
}
\author{Hina Tabassum, Ahmed Hamdi Sakr, and Ekram Hossain
\thanks{The work was supported by  the Natural Sciences and Engineering Research Council of Canada (NSERC).}
}

\maketitle

\thispagestyle{plain}
\pagestyle{plain}

\IEEEpeerreviewmaketitle

\begin{abstract}

Recent advancements in self-interference (SI) cancellation capability of low-power wireless devices motivate  in-band full-duplex (FD) wireless
backhauling in small cell networks (SCNs). In-band FD wireless backhauling allows the use of same frequency spectrum for the backhaul as well as access links of the small cells concurrently.  In this paper, using tools from stochastic geometry, we develop a framework to model the downlink rate coverage probability of a user in a given SCN with massive MIMO-enabled wireless backhauls. The considered SCN is composed of  a mixture of small cells that are configured in either in-band or out-of-band backhaul modes with a certain probability.  The performance of the user in the considered hierarchical network is limited by several sources of interference such as  the backhaul interference, small cell base station (SBS)-to-SBS interference and the SI. Moreover, due to the channel hardening effect in massive MIMO, the backhaul links experience long term channel effects only, whereas the access links  experience both the long term  and short term channel effects. Consequently, the developed framework is  flexible to  characterize different sources of interference while capturing the heterogeneity of the access and backhaul channels. In specific scenarios, the framework enables deriving closed-form coverage probability expressions. Under perfect backhaul coverage, the simplified expressions are utilized to optimize the proportion of in-band and out-of-band  small cells in the SCN in closed-form. Finally, few remedial solutions are proposed that can potentially mitigate the backhaul interference and in turn improve the performance of in-band FD wireless backhauling. Numerical results  investigate the scenarios in which in-band wireless backhauling is useful and  demonstrate that maintaining a  correct proportion of in-band and out-of-band FD small cells is crucial in wireless backhauled SCNs.
\end{abstract}

\section{Introduction}

Massive deployment of  small cells will be a key feature of the emerging 5G cellular networks~\cite{11}. Subsequently, efficient forwarding of the backhaul traffic of the small cells  will be a key challenge. By definition, the small cell backhaul connections are used to (i)~forward/receive the end-user (small cell user) data  to/from  the core network  and (ii)~exchange mutual information among different small cells. Typically, the backhaul of small cells  exploits either wired connectivity (e.g., optical fiber and DSL connections) or wireless connectivity (e.g., microwave or millimeter wave links). However, the prohibitively high cost of wired  connectivity,  directed transmission requirement of the microwave links,  and the poor penetration of the mmwave links make them less attractive backhaul solutions. To this end, leveraging the use of  radio access network (RAN) spectrum simultaneously for both access and backhaul links  has  emerged as a potential backhaul solution~\cite{uzma,1}.  Backhauling based on RAN spectrum  exhibits less-sensitivity to channel propagation effects, provide wider coverage due to non-LOS propagation, makes the reuse of existing hardware possible, and thus enable simpler operation and maintenance (O\&M). Nonetheless, the RAN spectrum availability is quite limited.

To enable efficient use in the RAN spectrum, full-duplex (FD) 
transmission  has been recently considered as a viable solution for 5G networks. FD transmission implies simultaneous transmission and reception of information in the same frequency band. FD transmission can be realized at base stations (BSs) through two different antenna configurations, i.e., shared and separated antenna configurations~\cite{liuband}.  The shared antenna configuration uses
a single antenna for simultaneous in-band
transmission and reception through a three-port circulator. 
On the other hand, the separated antenna configuration requires
separate antennas for transmission and reception. 
The gains of FD transmission are however limited by the 
overwhelming nature of  self-interference (SI), which is generated by the transmitter  to its own collocated receiver \cite{madu}.  Fortunately, with the recently developed antenna and digital baseband technologies, SI can be reduced close to the level of noise floor in low-power devices \cite{13}.  
As such, FD transmission in the access and backhaul links of small cells  enables reuse of RAN spectrum,  alleviates the need to procure spectrum for backhauling, and facilitates hardware implementation. 

\subsection{Background Work}

Several recent studies focus on developing backhaul interference management schemes~\cite{bennis,7} and backhaul delay minimizing solutions~\cite{tony}. In~\cite{bennis}, an interference management strategy is proposed for self-organized small cells. The small cell base stations (SBSs) operate like decode-and-forward relays for the macrocell users and forward their  uplink traffic to the macrocell base station (MBS) over heterogeneous backhauls. 
The problem is formulated as a non-cooperative game and  a reinforcement learning approach is used to find an equilibrium. In \cite{7}, a duplex and spectrum sharing scheme based on co-channel reverse time-division duplex (TDD) and dynamic soft frequency reuse (SFR), are proposed for backhaul interference management. An optimization  problem is formulated to allocate backhaul bandwidth and to optimize user association such that the network sum-rate is maximized. 
A tractable model is developed in \cite{tony} to characterize the  backhaul delay experienced by a typical user in the downlink considering both wired and wireless backhauls.   It is shown that deploying dense small cell networks may not be effective without a comparable investment in the backhaul network.  

Along another line of research, the performance gains of FD SBSs are recently 
analyzed for concurrent uplink and downlink transmissions in various research studies considering a single antenna \cite{16,ibfd,ibfd1,ibfd2,15} and multi-antenna \cite{17,ibfd3} transmissions. 
In \cite{16}, the feasibility conditions of FD operation are investigated followed by developing an uplink/downlink user scheduling scheme that maximizes the overall utility of all users.  
In \cite{ibfd}, a resource management scheme is developed that assigns downlink and uplink transmissions jointly and perform mode selection at each resource block while considering the gain of SI  cancellation. Users' transmit power levels are then determined such that the total utility sum is maximized.
Another optimization-based study considers maximizing the sum-rate performance by jointly optimizing subcarrier assignment and power allocation considering FD transmissions~\cite{ibfd1}. 

{In \cite{ibfd2}, considering Rayleigh fading channels and ideal backhauls for SBSs, a downlink throughput analysis of a $K$-tier network is conducted in which the SBSs operate in either a downlink half-duplex (HD)  or 
a bi-directional FD\footnote{Bi-directional FD mode allows both the transmitter and the receiver to transmit signals to each other at the same time and in the same frequency band.} mode.  It is shown that  all BSs  should operate in either  HD or FD mode to maximize network throughput.} Using Wyner model,  \cite{15} analyzes  the  network rate under  single-cell processing and cloud-RAN operation considering either HD
or  FD  BSs. Insights are extracted related to the  regimes  in  which  FD  BSs are  advantageous. In \cite{17}, the gains of using multiple-antennas for  HD  multiple-input-multiple-output (MIMO)  link  are compared to  the gain
achieved by utilizing  the antennas  for   FD  transmission. Conditions are analyzed in which using additional antennas for  FD transmission  is beneficial instead of high capacity HD MIMO link. 
For a single massive MIMO-enabled MBS and small cells deployed on a fixed distance from the MBS (ignoring the co-tier interference among SBSs),
a precoding method is designed in \cite{ibfd3}. The precoding method  enables complete rejection of backhaul interference received at a given user from the MBS to which the serving SBS of that user is associated with.

\subsection{Motivation and Contributions}
Although the feasibility  of  HD SBSs (referred as out-of-band FD SBSs in this paper)  has been investigated for wireless backhauls, there has not been any comprehensive study on the performance and feasibility of FD SBSs (referred as in-band FD SBSs in this paper) in wireless backhauled small cell networks. In particular, the use of full-duplex communication  in wireless backhauling and the performance gains have not been investigated. With this motivation, this paper provides a framework to understand the performance and significance of full-duplex communication in wireless backhaul networks.
While the in-band FD operation can ideally double the spectral efficiency in a link, the network-level gain of exploiting FD transmission in the wireless backhauls remains unclear due to the complicated interference environments, e.g., SI, co-tier, and cross-tier interferences at the backhaul and access links (as illustrated in Fig.~1). 
Due to the complicated interference environments in full-duplex backhaul transmission scenarios,  a theoretical framework  is required to characterize the diverse interference issues and to critically analyze the  scenarios in which in-band backhauling may be beneficial over out-of-band backhauling. In this context, the contributions of this paper are listed as follows:
\begin{itemize}
\item Using  tools from stochastic geometry, we model the performance of a massive MIMO-enabled wireless backhaul network that supports single-antenna small cells.  A hierarchical network structure is considered in which the massive MIMO-enabled wireless backhaul hubs (or connector nodes (CNs)) are deployed to provide simultaneous backhaul to multiple SBSs. Each SBS can be configured either in the in-band or out-of-band FD backhaul mode with a certain probability. 
{Note that the  in-band  FD backhaul mode leads to three-node full-duplex (TNFD) transmission which involves three nodes, i.e., the CN transmits to an SBS and the SBS transmits to a user\footnote{Interested readers are referred to \cite{madu} for the details of  TNFD mode of  transmission. In the TNFD mode, a typical user suffers from  backhaul interference (i.e., the interference received at a typical user from the CN  which is associated to its designated in-band FD SBS).}.} 

\item {The downlink rate coverage (which is a function of the rate coverage  in the access and backhaul links) of a small cell user is derived
considering both the in-band and out-of-band FD backhaul modes of a given SBS. Due to the channel hardening effect in massive MIMO, the backhaul links experience long-term channel effects only, whereas the access links  experience both the long term  and short term channel effects. Thus, unlike traditional Rayleigh fading assumption, the framework captures the heterogeneity of the access and backhaul links.}

\item Closed-form rate coverage expressions are then provided for specific scenarios. The simplified expressions, under the assumption of perfect backhaul coverage, are then utilized to customize the proportion of in-band and out-of-band FD SBSs in a small cell network in closed-form.

\item  A distributed backhaul interference-aware mode selection mechanism is then discussed to gain insights into selecting the proportion of in-band and out-of-band FD SBSs as a function of network parameters. 

\item {Due to backhaul interference, downlink transmission to a user by an SBS is directly affected by the transmission in the backhaul link to this SBS. We therefore  present few remedial solutions that can potentially mitigate the backhaul interference and quantify the enhancements in  the performance of in-band FD wireless backhauls.}
\end{itemize}
Numerical results demonstrate the usefulness of in-band FD wireless backhauls in scenarios with low SI, low transmit power and intensity of CNs, and higher intensity of SBSs. It is also shown that solely implementing the in-band or out-of-band FD backhaul solutions may not be useful. Instead, a system with the correct proportion of in-band and out-of-band FD SBSs  should be implemented.

\subsection{Paper Organization and Notations}

The rest of the paper is structured as follows. Section~II discusses the system model and assumptions, the channel model, and the massive MIMO-enabled wireless backhaul model. Section~III formulates the signal-to-interference-plus-noise ratio (SINR) model, performance metrics, and discusses some of the approximations considered throughout the paper. 
Section~IV derives the rate coverage  expressions for a given SBS considering both in-band and out-of-band FD backhaul modes. Section~V provides the simplified rate coverage  expressions and customizes the proportion of  in-band and out-of-band FD SBSs in the network. Backhaul interference management techniques are discussed in Section~VI. Section~VII provides numerical and simulation results followed by the concluding remarks in Section~VIII.

\vspace{1 mm}
\noindent\textbf{Notation}: 
$\mathrm{Gamma}(k_{(\cdot)},\theta_{(\cdot)})$ represents a Gamma distribution with shape parameter $k$, scale parameter $\theta$ and $(\cdot)$ displays the name of the random variable~(RV). $\Gamma(a)=\int_0^\infty x^{a-1} e^{-x} dx$ represents the Gamma function, ${\Gamma}_u (a;b)=\int_b^\infty x^{a-1} e^{-x} dx$ denotes the upper incomplete Gamma function, and ${\Gamma}_l(a;b)=\int_0^b x^{a-1} e^{-x} dx$ denotes the lower incomplete Gamma function~\cite{book}. $_2F_1[\cdot,\cdot,\cdot]$ denotes the Gauss Hypergeometric function and $\mathcal{B}(\cdot,\cdot,\cdot)$ denotes the incomplete Beta function. $f(\cdot)$ and $F(\cdot)$ denote the probability density function (PDF) and cumulative distribution function (CDF), respectively. Finally, $\mathbb{E}[\cdot]$ denotes the expectation operator.

\section{System Model and Assumptions}

\subsection{Network Deployment Model}

We consider a downlink two-tier cellular network with  small cells  connected to the core network through connector nodes (CNs). The CNs  provide wireless backhaul connections to the small cells. A CN is typically situated at a fiber point-of-presence or where high-capacity line-of-sight (LOS) microwave link is available to the core network. An existing MBS can be an example of such a CN which is connected to the core network by fiber-to-the-cell (FTTC) links. The locations of the CNs, SBSs, and users are modeled according to independent homogeneous Poisson Point Processes (PPPs) $\Phi^\prime_{\mathrm{c}}$, $\Phi^\prime_{\mathrm{s}}$, and $\Phi^\prime_{\mathrm{u}}$ with intensities $\lambda^\prime_c$,  $\lambda^\prime_s$, and $\lambda^\prime_u$, respectively. 
{Specifically, for a given PPP, the number of points and their locations are random and they follow  Poisson and uniform distributions, respectively.} {The abstraction model where the nodes are distributed using PPP  has been shown to be quite effective for system-level performance evaluation of cellular networks~(see \cite{elsawy2013stochastic} and references therein)\footnote{{Other modeling options may also be used, e.g., the locations of the CNs might be correlated with the SBSs. This can be taken into account by considering non-Poisson point processes. However, for analytical intractability, such a modeling option is not considered in this paper.}}.}

For the downlink transmission, each user (and SBS) associates to the SBS (and CN) that offers the strongest average received power.
{Since the users follow a PPP, therefore, the number of users in the system as well as the number of users associated per SBS are random. Several users associated to a given SBS  are assumed to be served at orthogonal frequency resource blocks.}

\begin{figure*}
\centering
\begin{minipage}{.5\textwidth}
  \centering
  \includegraphics[width=.9\linewidth]{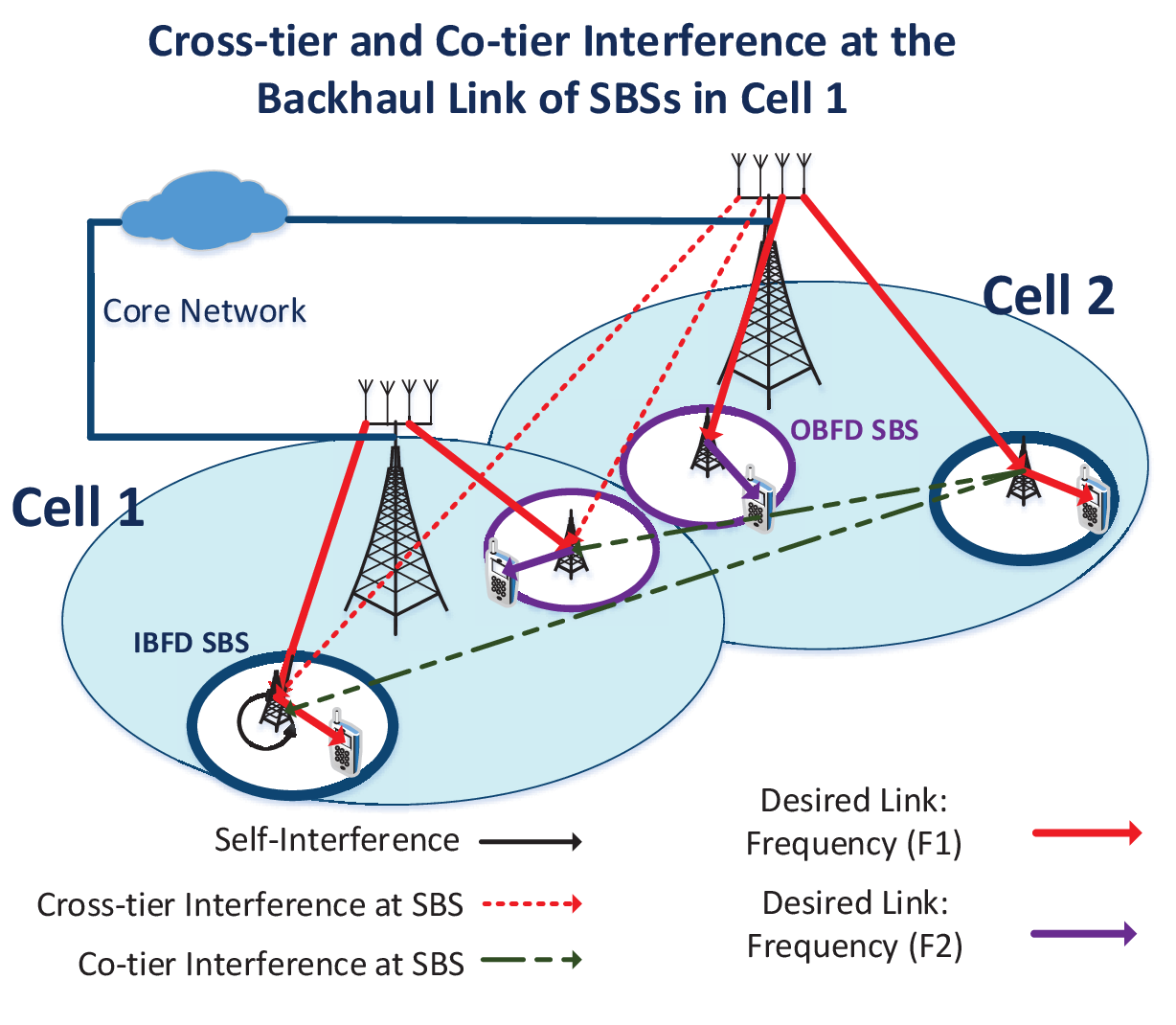}
\end{minipage}%
\begin{minipage}{.5\textwidth}
  \centering
  \includegraphics[width=.9\linewidth]{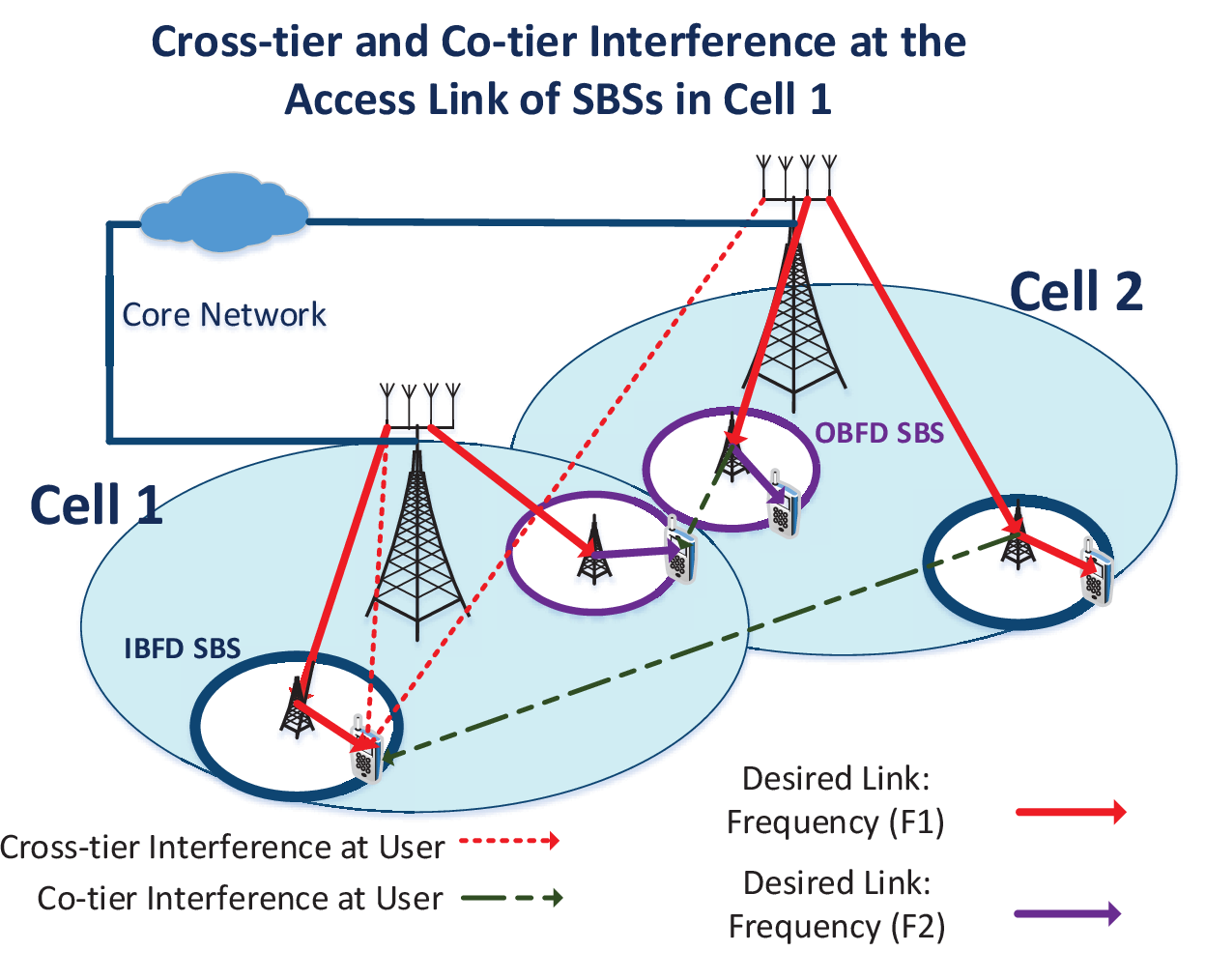}
\end{minipage}
\caption{Graphical illustration of the considered massive MIMO-enabled wireless backhaul network for both in-band and out-of-band small cells. Co-tier and cross-tier interferences experienced at the access and backhaul links are illustrated for in-band and out-of-band small cells located in cell~1.}
\end{figure*}

\subsection{Channel Model}
Throughout the paper, the propagation channels related to network elements of the same type are assumed to be independent and identically distributed (i.i.d.).  If a generic CN-to-SBS link is considered, e.g.,
from the CN $i$ to the SBS $k$, the channel parameters are identified by using the subscript $\{\mathrm{c_i, s_k}\}$. A similar notation holds for the channel parameters related to other network elements. The wireless channels are subject to  path-loss, shadowing, and fast-fading.

\subsubsection{Path-Loss}
Let ${r}_{X_i,Y_k}$ be the distance between two generic network elements $X_i$ and $Y_k$. Based on the downlink network model, we have $X_i \in \{\mathrm{c}_i, \mathrm{s}_i\}$ and $Y_k \in \{\mathrm{s}_k,  \mathrm{u}_k\}$ where $\mathrm{c}, \mathrm{s}$, and $\mathrm{u}$ denote the generic CN,  SBS, and user, respectively. The path-loss of this generic link is defined as ${r}^{-\beta}_{X,Y}$ (for ease of exposition, we exclude the subscripts of $X$ and $Y$) where $\beta > 2$ denotes the  path-loss exponent.

\subsubsection{Shadowing and Fading}
In addition to the distance-dependent path-loss, the generic link  is subject to both the shadowing $S_{X,Y}$ and  fast fading $F_{X,Y}$, unless stated otherwise. We model shadowing  with the log-normal distribution where
$\mu_{X,Y}$ and $\sigma_{X,Y}$ are  the mean and standard deviation of the shadowing channel power, respectively. Furthermore, the fading component of the wireless channel is modeled by the Nakagami-\emph{m} distribution. Nakagami-\emph{m}  is a generic fading distribution which includes Rayleigh distribution for $m=1$ (typically used for non-LOS conditions) as a special case and can well-approximate the Rician fading distribution for $1 \leq m\leq \infty$ (typically used for strong LOS conditions)\cite{KG}. 

{\bf Remark~1}~(Displacement Theorem~\cite{6658810}){\bf.} The displacement theorem can be used to deal with the shadow fading ${S}_{X,Y}$ as a random and independent transformation of a given homogeneous PPP $\Phi$ of density $\lambda$. In this case, the resulting point process is also a PPP with equivalent density $\lambda \mathbb{E}[{S}_{X,Y}^{\frac{2}{\beta}}]$. Applying this theorem to our case, we can handle the effect of any distribution for the shadow fading as long as the fractional moment $\mathbb{E}[{S}_{X,Y}^{\frac{2}{\beta}}]$ is finite. 
For log-normal shadowing,   the fractional moment can be given by using the definition of its moment generating function (MGF) as follows:
\begin{equation}
\mathbb{E}[S^{\frac{2}{\beta}}_{X,Y}]=\exp \left(2 \frac{\mu_{X,Y}}{\beta} +0.5 \left(2 \frac{\sigma_{X,Y}}{\beta}\right)^2 \right).
\end{equation}
As such,  ${\Phi}^\prime_c$ and ${\Phi}^\prime_s$ become $\Phi_c$ and $\Phi_s$ with densities  $\lambda_c  ={\lambda}^\prime_c \mathbb{E}[{S}_{X,Y}^{\frac{2}{\beta}}]$ and $\lambda_s  = {\lambda}^\prime_s \mathbb{E}[{S}_{X,Y}^{\frac{2}{\beta}}]$, respectively.
Thus, a generic useful  or interfering link gain given by $ {r}_{X,Y}^{-\beta} S_{X,Y} F_{X,Y}$, where $r$ follows a PPP $\Phi^\prime$, will be modeled  as $ r^{-\beta}_{X,Y}  F_{X,Y}$ where $r$ follows a PPP $\Phi$ throughout the paper.

\subsection{Backhaul Modes of Operation at SBSs}
Each SBS can operate in two modes for backhaul transmission, i.e.,
\begin{itemize}

\item {\bf In-Band Full-Duplex (IBFD) mode:} in which the access link and backhaul link transmissions are conducted  in the same frequency band (say $F_1$) of bandwidth $B$. Thus,  the total bandwidth usage of IBFD mode is $B$.
\item {\bf Out-of-Band Full-Duplex (OBFD) mode:} in which the access link and backhaul link transmissions are conducted  in different frequency bands $F_2$ and  $F_1$, respectively. Note that, in OBFD mode, if we calculate capacity by considering $F_1$ and $F_2$ each of bandwidth $B$, then the total bandwidth usage of OBFD mode becomes $2 B$.
This is an unfair setting from the perspective of bandwidth usage and in turn capacity calculation.

\end{itemize}
In order to have a fair comparison, we need to assume that each mode  (i.e., IBFD or OBFD)  can consume a total bandwidth of $B$ Hz only, i.e.,  the IBFD mode would use $B$ Hz for both access and backhaul link transmissions. On the other hand, the OBFD mode would have 
to use $0.5B$ Hz of $F_2$ for the access link and $0.5B$ Hz of $F_1$ for the backhaul link. 
Detailed throughput calculations are shown in Section~III.A.

For efficient antenna usage, in this paper, FD operation is achieved at the SBS using a shared antenna that separates the transmitting and receiving circuit chains through  an ideal circulator. 
The transmission mode selection for  small cells is modeled by independent Bernoulli RVs such that small cells
are configured in IBFD and OBFD mode with probability $q$ and $1-q$,
respectively, where $q$ is identical for all small cells. 
With the independent thinning of $\Phi_{\mathrm{s}}$, we can represent the SBSs in IBFD and OBFD modes as two independent PPPs $\Phi_{\mathrm{sI}}$ and $\Phi_{\mathrm{sO}}$ with density $q \lambda_s$ and  $(1-q) \lambda_s$, respectively.

\subsection{Massive MIMO-enabled Backhaul Model}

Each CN is equipped with $M$ antennas that serves the backhaul links of a maximum of $\mathcal{S}$ single-antenna SBSs. {Each SBS serves one user at a time and  each user device is equipped with one antenna.}
We refer to the massive MIMO regime as the case
where $1 \ll \mathcal{S} \ll M$.
{During the training phase, each SBS sends a pre-assigned orthogonal pilot sequence to the CN which is estimated perfectly by the CN and the pilot sequence is not used by any other CN (i.e., no pilot contamination is assumed)\footnote{{Some guidelines will be provided in Section~VII on how to incorporate the effect of interference due to pilot contamination.}}.} The channel estimation is facilitated by considering  time division duplexing (TDD)  at CNs such that the channel reciprocity is guaranteed. The maximum number of downlink backhaul data streams $\mathcal{S}$ per CN depends on the dimension of the uplink pilot field, which in turn determines the number of  channel vectors that can be estimated and for which the downlink  precoder can be calculated. 

Each CN serves its SBSs by using  linear zero-forcing beamforming (LZFBF) with equal power per backhaul data stream. As such, in the massive MIMO 
regime, the effects of Gaussian noise and uncorrelated intra-cell interference disappear. {In this paper, massive MIMO is utilized at CNs to facilitate  simultaneous backhaul transmissions to several SBSs on the same frequency channel. In particular, with the aid of massive MIMO and LZFBF, the interference among multiple backhaul streams of a CN is mitigated.
However, due to this interference mitigation,  the massive MIMO with LZFBF could possibly be more beneficial for IBFD SBSs as their both access/backhaul links are vulnerable to the backhaul transmissions.  This is in contrast to OBFD SBSs where backhaul interference effects only the backhaul links (as can also be depicted from Fig.~1). }
The  rates of backhaul links to different SBSs thus tend to concentrate on deterministic limits that are calculated using tools from asymptotic random matrix theory. In this paper, we use the following  formula for the backhaul data rate (in bps) in the link between a CN and a given SBS \cite{mimo2}:
\begin{equation}\label{backhaul}
R_{b}=\alpha B \mathrm{log}_2\left(1+\frac{M-\mathrm{min}(\mathcal{N}_s, \mathcal{S})+1}{\mathrm{min}(\mathcal{N}_s, \mathcal{S})} \mathrm{SIR}_b\right),
\end{equation}
where $\mathcal{N}_s$ denotes the number of SBSs associated with the CN, $\mathrm{min} (\mathcal{N}_s, \mathcal{S})$ is the number of SBSs served by the CN at a time, the factor in the numerator ${M-\mathrm{min}(\mathcal{N}_s, \mathcal{S})+1}$ is the massive MIMO gain, the factor ${\mathrm{min}(\mathcal{N}_s, \mathcal{S})}$ in the denominator is due to dividing the CN's transmit power equally per backhaul stream in the useful link, and $\mathrm{SIR}_b$ denotes the signal-to-interference ratio (SIR)  at the  SBS (with some abuse of the notation due to the multiplying factor $\frac{M-\mathrm{min}(\mathcal{N}_s, \mathcal{S})+1}{\mathrm{min}(\mathcal{N}_s, \mathcal{S})}$), {and $\alpha$ is the {\em backhaul access probability} of a given  SBS, which is defined as  the probability that an SBS is served by a CN over a backhaul link}. Without loss of generality, we ignore the impact of the time required for channel estimation in \eqref{backhaul}; however, it can be incorporated by multiplying $R_{b}$ with the ratio of time spent sending data to the total time frame~\cite{marzetta2010noncooperative}.

\subsection{Load of CNs and Backhaul Access Probability of  SBSs}
Note that the number of SBSs $\mathcal{N}_s$ associated to a given CN using 
the strongest signal association policy is a RV and can exceed the supported number of backhaul streams $\mathcal{S}$.  The probability mass function (PMF) of the number of SBSs served by a generic CN can be given as follows \cite{SakrD2D}:
\begin{equation}\label{pmf}
\mathbb{P}(\mathcal{N}_s=n)= \frac{b^b \Gamma(n+b) (\mathbb{E}[\mathcal{N}_s])^n}
{\Gamma(n+1)\Gamma(b) (b+\mathbb{E}[\mathcal{N}_s])^{n+b}},
\end{equation}
where $\mathbb{E}[\mathcal{N}_s]=\lambda_s/\lambda_c$
is  the average  number  of  SBSs  per  CN.  This  expression  is
derived  by  approximating  the  area  of  a  Voronoi  cell  by  a
gamma-distributed  RV  with  shape  parameter $b=3.575$
and scale parameter $\frac{1}{b \lambda_c}$. Note that this expression
is  valid  only  when  the  SBSs  are  assumed  to  be
spatially  distributed  according  to  an  independent  PPP.

In case when  all SBSs associated to a given CN cannot  be served at the same time, the CN  randomly picks a set of $\mathcal{S}$ SBSs. The backhaul access probability of an SBS can thus be derived as follows:
\begin{align}\label{alpha}
\alpha=
\begin{cases}
1, & \mathcal{N}_s \leq \mathcal{S}\\
\frac{{{\mathcal{N}_s-1} \choose \mathcal{S}-1}}{{{\mathcal{N}_s} \choose \mathcal{S}}}=\frac{\mathcal{S}}{\mathcal{N}_s}, & \mathcal{N}_s > \mathcal{S}
\end{cases}
=
\min\left(1,\frac{\mathcal{S}}{\mathcal{N}_s}\right).
\end{align}

\section{Performance Metrics and SINR Model of IBFD and OBFD Transmission Modes}

In this section, we define the performance metrics of interest, characterize the different sources of interference, and formulate the SINR  at the access and backhaul links of the  SBSs in both the IBFD and OBFD modes of backhaul operation. Some relevant remarks regarding the intensity of interfering sources and approximations are also highlighted which will be applied for the analytical evaluation of the rate coverage probability expressions later in Section~IV.

\subsection{Performance Metrics}

From Section~II.C, the instantaneous downlink rates (in bps) at the access link of a given SBS  in the IBFD and OBFD modes  can be defined, respectively, as:
\begin{align}
{\mathcal{\tilde{R}}_{\mathrm{a,I}}}=& {B}
\mathrm{log}_2(1+\mathrm{SINR}_{\mathrm{a,I}}),
\nonumber\\
{\mathcal{\tilde{R}}_{\mathrm{a,O}}}=&\frac{{B}}{2} \mathrm{log}_2(1+\mathrm{SINR}_{\mathrm{a,O}}).
\end{align}

Similarly, the instantaneous downlink rates (in bps) at the backhaul link of a given SBS  in the IBFD and OBFD modes  are defined using \eqref{backhaul}, respectively, as follows:
\begin{align}
{\mathcal{\tilde{R}}_{\mathrm{b,I}}}= &{\alpha} B \mathrm{log}_2
\left(1+\frac{M-\mathrm{min}(\mathcal{N}_s, 
\mathcal{S})+1}{\mathrm{min}(\mathcal{N}_s, \mathcal{S})}\:\: \mathrm{SIR}_{\mathrm{b,I}}\right),
\nonumber\\
{\mathcal{\tilde{R}}_{\mathrm{b,O}}}=&\frac{{\alpha} {B}}{2} \mathrm{log}_2
\left(1+\frac{M-\mathrm{min}(\mathcal{N}_s, \mathcal{S})+1}{\mathrm{min}(\mathcal{N}_s, \mathcal{S})}\:\: \mathrm{SIR}_{\mathrm{b,O}}\right).
\end{align}

Therefore, the {\bf normalized rate (in bps/Hz)} (i.e., the spectral efficiency) for IBFD and OBFD modes can be obtained, respectively, by dividing all aforementioned rate equations with the bandwidth $B$, i.e.,
\begin{align}\label{shannon}
\mathcal{R}_{\mathrm{a,I}}=& 
\mathrm{log}_2(1+\mathrm{SINR}_{\mathrm{a,I}}),
\nonumber\\
\mathcal{R}_{\mathrm{a,O}}=&\frac{1}{2} \mathrm{log}_2(1+\mathrm{SINR}_{\mathrm{a,O}}).
\end{align}

Similarly, the downlink rates (in bps/Hz) at the backhaul link of a given SBS  in the IBFD and OBFD modes  are defined, respectively, as follows:
\begin{align}\label{backhaul1}
\mathcal{R}_{\mathrm{b,I}}= &{\alpha} \mathrm{log}_2
\left(1+\frac{M-\mathrm{min}(\mathcal{N}_s, 
\mathcal{S})+1}{\mathrm{min}(\mathcal{N}_s, \mathcal{S})}\:\: \mathrm{SIR}_{\mathrm{b,I}}\right),
\nonumber\\
\mathcal{R}_{\mathrm{b,O}}=&\frac{{\alpha}}{2} \mathrm{log}_2
\left(1+\frac{M-\mathrm{min}(\mathcal{N}_s, \mathcal{S})+1}{\mathrm{min}(\mathcal{N}_s, \mathcal{S})}\:\: \mathrm{SIR}_{\mathrm{b,O}}\right).
\end{align}

Without loss of generality,  we use normalized rate (or spectral efficiency) for numerical evaluation and analysis throughout the paper.

\subsubsection{Downlink rate coverage of an SBS in IBFD mode}
Now let us define the rate coverage probability $\mathcal{C}_I$ of a given SBS operating in IBFD mode as  the probability that the downlink user rate is higher than a required  target rate $R_{\mathrm{th}}$. To achieve $R_{\mathrm{th}}$  in IBFD mode, both $\mathcal{R}_{\mathrm{b,I}}$ and $\mathcal{R}_{\mathrm{a,I}}$  must be at least  $R_{\mathrm{th}}$. In other words,  $\mathrm{SINR}_{\mathrm{a,I}}$ must be greater than a prescribed threshold $\gamma_{\mathrm{a,I}}$ to achieve a target rate of $R_{\mathrm{th}}$ in the access link, where $\gamma_{\mathrm{a,I}}$ can be given using \eqref{shannon} as follows:
\begin{equation}
\gamma_{\mathrm{a,I}}= 2^{{{R_{\mathrm{th}}}}}-1.
\end{equation}
On the other hand,
$\mathrm{SIR}_{\mathrm{b,I}}$ must be greater than another prescribed threshold $\gamma_{\mathrm{b,I}}$ to achieve a target rate of $R_{\mathrm{th}}$ in the backhaul link. Using \eqref{backhaul1},  $\gamma_{\mathrm{b,I}}$ can be defined as follows:
\begin{equation}\label{gammabI}
\gamma_{\mathrm{b,I}}= \frac{\mathrm{min}(\mathcal{N}_s,\mathcal{S})}{M-\mathrm{min}(\mathcal{N}_s,\mathcal{S})+1}\left(2^{{\frac{R_{\mathrm{th}}}{\alpha}}}-1\right).
\end{equation}
The coverage probability $\mathcal{C}_{\mathrm{I}}$ of an SBS in the IBFD mode  is  a function of the 
${\rm SINR}$ at both the access link and the backhaul link, and the ${\rm SINR}$ outage probability   
$\mathcal{C}_{\mathrm{I}}$ is defined as follows:
\begin{equation}\label{ci}
\mathcal{C}_{\mathrm{I}}=\mathbb{P}(\mathrm{SIR}_{\mathrm{b,I}} > \gamma_{\mathrm{b,I}}) \; \mathbb{P}(\mathrm{SINR}_{\mathrm{a,I}} > \gamma_{\mathrm{a,I}}).
\end{equation}

\subsubsection{Downlink rate coverage of an SBS in OBFD mode}
The rate coverage probability of an SBS in the OBFD mode is given as
\begin{equation}\label{co}
\mathcal{C}_{\mathrm{O}}=\mathbb{P}(\mathrm{SIR}_{\mathrm{b,O}} > \gamma_{\mathrm{b,O}}) \; \mathbb{P}(\mathrm{SINR}_{\mathrm{a,O}} > \gamma_{\mathrm{a,O}}),
\end{equation}
where $
\gamma_{\mathrm{a,O}}= 2^{2\frac{R_{\mathrm{th}}}{{B}}}-1
$  from (\ref{shannon}). To achieve $R_{\mathrm{th}}$  in the backhaul link,  $\gamma_{\mathrm{b,O}}$ can be defined from \eqref{backhaul1} as follows:
\begin{equation}
\gamma_{\mathrm{b,O}}=\frac{\mathrm{min}(\mathcal{N}_s,\mathcal{S})}{M-\mathrm{min}(\mathcal{N}_s,\mathcal{S})+1}\left(2^{\frac{2 R_{\mathrm{th}}}{\alpha}}-1\right).
\end{equation}

\subsubsection{Downlink rate coverage  of a typical user}
The overall rate coverage probability of a typical user depends on 
the probability of a given SBS to operate in  OBFD or IBFD mode and can thus be given as
\begin{equation}
\mathcal{C}_{u}=q \mathcal{C}_{\mathrm{I}} +(1-q)\mathcal{C}_{\mathrm{O}}.
\end{equation}

\subsection{SINR Model for IBFD Mode}
\subsubsection{Access link} The SINR of a generic user  associated with an SBS in the IBFD mode   can be defined as follows:
\begin{equation}\label{SINRa}
\mathrm{SINR}_{\mathrm{a,I}}=\frac{P_s \; F_{\mathrm{s},\mathrm{u}} \; r_{\mathrm{s},\mathrm{u}}^{-\beta} }{I_{\mathrm{s},\mathrm{u}} + I_{\mathrm{c},\mathrm{u}} +N_0},
\end{equation}
where $P_s$ denotes the transmit power of a generic SBS,  $r_{\mathrm{s},\mathrm{u}}$ and $F_{\mathrm{s},\mathrm{u}}$ denote the distance  and the Gamma distributed fading between a user and its serving SBS, respectively, and $N_0$ is the thermal noise power.  Since  $\Phi_{\mathrm{s}}$ is a stationary PPP, the PDF of the distance $r_{\mathrm{s},\mathrm{u}}$ is given by the Rayleigh distribution, i.e., 
$
f_{{r_{\mathrm{s},\mathrm{u}}}}({r})= 2 \pi  \lambda_s {r} e^{-\pi  \lambda_s {r}^2}$. 

{In the IBFD mode, a given user experiences  interference from all  SBSs in IBFD mode which receive backhaul data from their corresponding CN.} Thus, $I_{\mathrm{s},\mathrm{u}}$ can be modeled as follows:
\begin{equation}
I_{\mathrm{s},\mathrm{u}}=\sum_{x\in\bar{\Phi}_{\mathrm{sI}}\setminus \mathrm{s}} P_s \; F_{\mathrm{x},\mathrm{u}} \;r_{\mathrm{x},\mathrm{u}}^{-\beta},
\end{equation}
where $\bar{\Phi}_{\mathrm{sI}}$ denotes the PPP of successfully backhauled SBSs in IBFD mode with intensity $\bar{\lambda}_{\mathrm{sI}}$ and can be given using the {\bf mean load approximation} and {\bf Remark~2}  detailed below.

\vspace{1mm}
{\bf Mean load approximation:} For analytical tractability,  we approximate the number of SBSs $\mathcal{N}_s$ associated to a given  CN  by its average value, i.e., $\mathbb{E}[\mathcal{N}_s]=\lambda_s/\lambda_c$. 
As such, 
the average number of  SBSs in IBFD mode served per CN can be given by $q \min\left(\lambda_s/\lambda_c,\mathcal{S}\right)$. 

\vspace{1mm}
{\bf Remark~2}~(Intensity of Interfering SBSs in IBFD Mode){\bf.} Given the average number of  SBSs in IBFD mode served per CN, the intensity $\bar{\lambda}_{\mathrm{sI}}$ of the interfering SBSs in IBFD mode can be  defined  as $q \min\left(\lambda_s/\lambda_c,\mathcal{S}\right) \lambda_c$. 
The impact of the  {\bf mean load approximation} will be demonstrated through numerical results in Section~VII.

The interference at a user due to backhaul transmissions can be modeled as follows: 
{\begin{equation}\label{Icu}
I_{\mathrm{c},\mathrm{u}}  = \sum_{y\in {\Phi}_{c}} {P}_c \; r_{\mathrm{y},\mathrm{u}}^{-\beta}, 
\end{equation}}
Note that the interference at a user will be received from all streams of a given CN thus the total interfering power from a CN is  ${P}_c$ and there is no short-term fading factor in the interference experienced from  the CNs due to the channel hardening effect.

\subsubsection{Backhaul link} 
The received SIR on the backhaul link of a given SBS in IBFD mode is given as:
\begin{equation}\label{eq:SINRbI}
\mathrm{SIR}_{\mathrm{b,I}} = \frac{{{P}_c} \; r_{\mathrm{c},\mathrm{s}}^{-\beta} }{I_{\mathrm{s,s}} + I_{\mathrm{c,s}} + I_{\mathrm{SI}}}, 
\end{equation}
where $r_{\mathrm{c},\mathrm{s}}$ denotes the  distance between the SBS and its serving CN $c$. Note that the impact of equal power allocation per backhaul stream (i.e., the denominator of CN's transmit power $\frac{P_c}{\mathrm{min}({\mathcal{N}_s},\mathcal{S})}$) has already been incorporated in the desired SIR threshold $\gamma_{\mathrm{b,I}}$ (see \eqref{backhaul} and \eqref{gammabI}). Therefore it is not considered in \eqref{eq:SINRbI}.
Since  $\Phi_c$ is stationary PPP, the
PDF of the distance $r_{\mathrm{c},\mathrm{s}}$
between each CN and its designated SBS  is given by
the Rayleigh distribution, i.e.,
$
f_{{r_{\mathrm{c},\mathrm{s}}}}({r})= 2 \pi \lambda_c {r} e^{-\pi \lambda_c {r}^2}
$.

Since the  SI incurred at a given SBS in IBFD mode depends on its own transmit power, we can define the residual SI power after performing SI cancellation as follows:
\begin{equation}
I_{\mathrm{SI}}=\frac{P_s}{\xi},
\end{equation}
where $\xi$ represents the SI cancellation capability of the SBS. Note that $\xi$ depends on the nature of the SI cancellation algorithms. For ease of exposition, we  consider $\xi$ as a constant value  in  this  paper. 

In the backhaul, a given  SBS in the IBFD mode experiences interference from all other SBSs operating in IBFD mode receiving their backhaul data from their corresponding CN. Thus $I_{{\mathrm{s}, \mathrm{s}}}$ can be modeled as 
$
I_{{\mathrm{s}, \mathrm{s}}}=\sum_{x \in \bar{\Phi}_{\mathrm{sI}}} P_s \; F_{\mathrm{x},\mathrm{s}} \; r_{\mathrm{x},\mathrm{s}}^{-\beta},
$
where $r_{\mathrm{x},\mathrm{s}}$  and $F_{\mathrm{x},\mathrm{s}}$ represent the  distance and fast fading channel between the two  SBSs operating in IBFD mode, respectively. Note that $\bar{\Phi}_{\mathrm{sI}}$ is defined by $\bar{\lambda}_{\mathrm{sI}}$ given in {\bf Remark~2}.
Finally, the interference received at an SBS from neighboring CNs  can be modeled as {
$
I_{{\mathrm{c}, \mathrm{s}}} = \sum_{y\in {\Phi}_{c}\setminus \mathrm{c}} {P}_c r_{{\mathrm{y}, \mathrm{s}}}^{-\beta}.
$}

\subsection{SINR Model for OBFD SBS}
\subsubsection{Access link}
The SINR received at a generic user  associated to an SBS $s$ operating in OBFD mode   can be defined as follows:
\begin{equation}
\mathrm{SINR}_{\mathrm{a,O}}=\frac{P_s \; F_{\mathrm{s},\mathrm{u}} \; r_{\mathrm{s},\mathrm{u}}^{-\beta} }{\hat{I}_{\mathrm{s},\mathrm{u}}+N_0},
\end{equation}
where {$\hat{I}_{\mathrm{s},\mathrm{u}}$ is the interference experienced by the user from all other SBSs operating in OBFD mode which are receiving their backhaul data from their corresponding CN}, i.e., 
$
\hat{I}_{\mathrm{s},\mathrm{u}}=\sum_{x\in\bar{\Phi}_{\mathrm{sO}}\setminus s} P_s F_{\mathrm{x,u}} r_{\mathrm{x,u}}^{-\beta}.
$
Here $\bar{\Phi}_{\mathrm{sO}}$ denotes the PPP of successfully backhauled SBSs in OBFD mode with intensity $\bar{\lambda}_{\mathrm{sO}}$ and can be obtained by using the {\bf mean load approximation}  as detailed below in {\bf Remark~3}.

\vspace{1mm}
{\bf Remark~3}~(Intensity of Interfering SBSs in OBFD Mode){\bf.}
Following similar steps as in {\bf Remark~2}, the intensity $\bar{\lambda}_{\mathrm{sO}}$ of the interfering SBSs in OBFD mode can be defined  as $(1-q) \min\left(\lambda_s/\lambda_c,\mathcal{S}\right) \lambda_c$. 

\subsubsection{Backhaul link}
The received SINR on the backhaul link of an SBS operating in OBFD mode  can be modeled as follows:
\begin{equation}\label{eq:SINR}
\mathrm{SINR}_{\mathrm{b,O}} = \frac{{{P}_c} \; r_{\mathrm{c},\mathrm{s}}^{-\beta} }{I_{\mathrm{s,s}} + I_{\mathrm{c,s}}}, 
\end{equation}
where
$I_{{\mathrm{s}, \mathrm{s}}}$ can be given as
$
I_{{\mathrm{s}, \mathrm{s}}}=\sum_{x \in \bar{\Phi}_{\mathrm{sI}}} P_s \; F_{\mathrm{x},\mathrm{s}} \; r_{\mathrm{x},\mathrm{s}}^{-\beta}
$
and
$
I_{{\mathrm{c}, \mathrm{s}}} = \sum_{y\in {\Phi}_{c}\setminus \mathrm{c}} {P}_c r_{{\mathrm{y}, \mathrm{s}}}^{-\beta}.
$

\vspace{1mm}
{\bf Remark~4}~(Backhaul Rate Coverage){\bf.}
Since $\mathcal{N}_s$ is a RV in the desired backhaul link threshold $\gamma_{\mathrm{b,I}}$,  the rate coverage of a given SBS in the IBFD mode can be calculated as follows:
\begin{multline}
\mathcal{C}_{\mathrm{I}}= \mathbb{P}(\mathrm{SINR}_{\mathrm{a,I}} > \gamma_{\mathrm{a,I}})
\\
\times
\mathbb{E}_{\mathcal{N}_s}\left[\mathbb{P}\left(\frac{{{P}_c} \; r_{\mathrm{c},\mathrm{s}}^{-\beta} }{I_{\mathrm{s,s}} + I_{\mathrm{c,s}} + I_{\mathrm{SI}}} > 
{ \gamma_{\mathrm{b,I}}}
\right)\right],
\end{multline}
where $\mathcal{N}_s \geq 1$.
The same method is applicable to the rate coverage of the SBS  in OBFD mode.

\section{Downlink Rate Coverage Analysis of IBFD/OBFD SBSs}

In this section, we  first derive the Laplace transforms of the interferences and then the rate coverage at the backhaul and access links of  a generic SBS considering both the  IBFD and OBFD modes with Gamma-distributed fading channel powers at the access links. 
The steps of the analytical approach are:
\begin{itemize}
\item Considering the IBFD mode, derive the Laplace transforms of the co-tier and cross-tier interferences received at the user, i.e., $\mathcal{L}_{I_{\mathrm{s},\mathrm{u}}}\left( t \right)$ and $\mathcal{L}_{I_{\mathrm{c},\mathrm{u}}}\left( t \right)$, respectively.
\item Derive the rate coverage at the access link of an SBS in IBFD mode. The rate coverage at the access link in OBFD mode is given as a special case of the IBFD mode.
\item Considering the IBFD mode, derive the Laplace transforms of the co-tier and cross-tier interferences received at the SBS, i.e., $\mathcal{L}_{I_{\mathrm{c},\mathrm{s}}}\left( t \right)$ and $\mathcal{L}_{I_{\mathrm{s},\mathrm{s}}}\left( t \right)$, respectively.
\item Derive the rate coverage at the backhaul link of an SBS in IBFD mode. The backhaul rate coverage  in OBFD mode is given as a special case of the IBFD mode.
\item Finally, the total rate coverage probability of a user in IBFD and OBFD mode, i.e.,  $\mathcal{C}_I$ and $\mathcal{C}_O$ are obtained using \eqref{ci} and \eqref{co}, respectively.
\end{itemize}

\subsection{Rate Coverage Analysis: Access Links}

To derive the rate coverage probability at the access link of a given SBS operating in IBFD mode, we first derive the Laplace transforms of the interferences $I_{\mathrm{s},\mathrm{u}}$ and $I_{\mathrm{c},\mathrm{u}}$ in the following.

Using the definition of the Laplace transform, the Laplace transform of the interference $I_{\mathrm{s},\mathrm{u}}$ can be derived as follows:
\begin{align}\label{eq:Lap_Isu}
&\mathcal{L}_{I_{\mathrm{s},\mathrm{u}}}\left( t \right) = \mathbb{E}_{I_{\mathrm{s},\mathrm{u}}} \left[ \exp \left[ -t I_{\mathrm{s},\mathrm{u}} \right] \right], 
\nonumber\\
&\stackrel{(a)}{=} \mathbb{E}_{\bar{\Phi}_{\mathrm{sI}},\{F_{\mathrm{x},\mathrm{u}}\}} \left[\prod_{x\in\bar{\Phi}_{\mathrm{sI}}\setminus \mathrm{s}} \exp \left[ -t P_s  
\;r_{\mathrm{x},\mathrm{u}}^{-\beta}
\;F_{\mathrm{x},\mathrm{u}}
\right] 
\right], 
\nonumber\\&
\stackrel{(b)}{=} \mathbb{E}_{\bar{\Phi}_{\mathrm{sI}}} 
\left[
\prod_{x \in \bar{\Phi}_{\mathrm{sI}} \setminus s} 
(1+t P_s  
\;r_{\mathrm{x},\mathrm{u}}^{-\beta}
\;\theta_{F_{\mathrm{x},\mathrm{u}}})^{-k_{F_{\mathrm{x},\mathrm{u}}} }
\right],
\nonumber\\
&\stackrel{(c)}{=} 
\exp \left[ - 2\pi \bar{\lambda}_{\mathrm{sI}} 
\int_{r_{\mathrm{s},\mathrm{u}}}^\infty \left(1-
\frac{1}{
{(1+\frac{t P_s  
\;\theta_{F_{\mathrm{x},\mathrm{u}}}}{\;r_{\mathrm{x},\mathrm{u}}^{\beta}})^{k_{F_{\mathrm{x},\mathrm{u}}}}}} \right)
r_{\mathrm{x},\mathrm{u}} d r_{\mathrm{x},\mathrm{u}} 
\right],
\nonumber\\
&\stackrel{(d)}{=} 
e^{ -\pi \bar{\lambda}_{\mathrm{sI}} r^2_{\mathrm{s},\mathrm{u}} 
\left(
\:_2F_1\left[k_{F_\mathrm{x,u}},-\frac{2}{\beta},
 1-\frac{2}{\beta},  -\frac{t  P_s  
\;\theta_{F_{\mathrm{x},\mathrm{u}}}}{ r_{\mathrm{s},\mathrm{u}}^{\beta}}\right]
-1
\right)
},
\end{align} 
where (a) follows from the independence of the interfering links, (b) follows from the definition of MGF of a Gamma RV, (c) follows  the probability generating functional (PGFL) of PPP,  and (d) follows from solving the integral  by making substitution $z = ({r_{\mathrm{s,u}}}/{r_{\mathrm{x,u}}})^\beta$ and performing algebraic manipulations. 

Similarly, following \eqref{eq:Lap_Isu}, the Laplace transform of $I_{\mathrm{c,u}}$ can be expressed as follows:
\begin{align}\label{eq:Lap_Icu}
\mathcal{L}_{I_{\mathrm{c,u}}}\left( t \right) &= 
\mathbb{E}_{{\Phi}_{\mathrm{c}}} 
\left[
\prod_{y \in {\Phi}_{\mathrm{c}}} 
\exp \left[ -t {P}_c  r_{\mathrm{y},\mathrm{u}}^{-\beta} \right] \right],
\nonumber\\
&\stackrel{(a)}{=} \exp \left[ - 2\pi {\lambda}_c \int_0^\infty \left(1 - \exp \left[ -t {P}_c  r_{\mathrm{y},\mathrm{u}}^{-\beta} \right] \right) r_{\mathrm{y},\mathrm{u}} \text{d}r_{\mathrm{y},\mathrm{u}}\right],
\nonumber \\
&\stackrel{(b)}{=} \exp \left[ - \pi {\lambda}_c (t {P}_c)^{\frac{2}{\beta}} \Gamma\left(1-\frac{2}{\beta}\right)\right],
\end{align}
where (a) follows from the PGFL of a PPP and (b) follows from solving the integral in (a). Note that the integral in \eqref{eq:Lap_Icu} has a  lower limit of  zero as the nearest CN of a user can  be arbitrarily close to the user.
 
The rate coverage at the access link of an SBS in IBFD mode can then be formulated as follows.
\begin{lemma}[Access Link Rate Coverage of an SBS in IBFD Mode: Gamma Fading Channels]
The coverage probability at the access link of an IBFD SBS can be derived in Gamma fading channels as 
\begin{align}\label{theorem}
&\mathbb{P}(\mathrm{SINR}_{\mathrm{a,I}}>\gamma_{\mathrm{a,I}})
=
\mathbb{P}\left(\frac{P_s \; F_{\mathrm{s},\mathrm{u}} \; r_{\mathrm{s},\mathrm{u}}^{-\beta} }{\underset
{I_{\mathrm{agg}}}
{\underbrace{I_{\mathrm{s},\mathrm{u}} + I_{\mathrm{c},\mathrm{u}} } } +N_0}>\gamma_{\mathrm{a,I}}\right),
\nonumber\\&
{=
2 \pi  \lambda_s
\int_0^\infty F_{{I_{\mathrm{agg}}}} \left( 
\frac{P_s \; F_{\mathrm{s},\mathrm{u}} \; }{\gamma_{\mathrm{a,I}} r_{\mathrm{s,u}}^{\beta}}  -N_0\right)r_{\mathrm{s,u}} e^{-\pi  \lambda_s {r_{\mathrm{s,u}}}^2}
\text{d}r_{\mathrm{s,u}}.}
\end{align}
Since fading in the desired channel follows a Gamma distribution, we resort to apply the Gil-Pelaez inversion theorem to evaluate the rate coverage probability. The CDF of the aggregate interference $F_{I_{\mathrm{agg}}} (\cdot)$ can be derived  as detailed below:
\begin{align}
&F_{I_{\mathrm{agg}}}\left( \frac{P_s \; F_{\mathrm{s},\mathrm{u}} }{r_{\mathrm{s},\mathrm{u}}^{\beta}\gamma_{\mathrm{a,I}}}-N_0 \right) 
\nonumber\\
&
\stackrel{(a)}{=} 
\frac{1}{2} -  \frac{1}{\pi}\int_0^\infty \mathrm{Im} \left[ \mathcal{L}_{I_{\mathrm{agg}}}\left( -j w \right)  e^{-j w 
\left( \frac{P_s \; F_{\mathrm{s},\mathrm{u}} }{r_{\mathrm{s},\mathrm{u}}^{\beta}\gamma_{\mathrm{a,I}}}-N_0 \right)} \right] \frac{\text{d} w}{w},\nonumber\\
&
\stackrel{(b)}{=} 
\frac{1}{2} -  \frac{1}{\pi}\int_0^\infty \mathrm{Im} \left[ \mathcal{L}_{I_{\mathrm{c,u}}}\left( -jw \right)
\mathcal{L}_{I_{\mathrm{s,u}}}\left( -jw \right)
\right.\nonumber\\&\left.
\:\:\:\:\:\:\:\:\:\:\:\:\:\:\:\:\:\:\:\:\:
\times
\mathcal{L}_{F_{\mathrm{s},\mathrm{u}}}\left(j \frac{P_s w }{r_{\mathrm{s},\mathrm{u}}^{\beta}\gamma_{\mathrm{a,I}}} \right)
e^{j w N_0 } \right] \frac{\text{d} w}{w},
\end{align}
\end{lemma}
where $\mathrm{Im}(\cdot)$ represents the imaginary part of the argument.  Using  \eqref{theorem}, (a) follows from the application of the  Gil-Pelaez inversion theorem and  (b) follows from the independence of $I_{\mathrm{s,u}}$ and $I_{\mathrm{c,u}}$ and applying the definition of the Laplace transform of $F_{\mathrm{s},\mathrm{u}}$ where $\mathcal{L}_{F_{\mathrm{s},\mathrm{u}}}(jw)$ can be given as follows:
\begin{equation}
\mathcal{L}_{F_{\mathrm{s},\mathrm{u}}}\left(j \frac{P_s w }{r_{\mathrm{s},\mathrm{u}}^{\beta}\gamma_{\mathrm{a,I}}} \right)=
\left(1+ j \frac{P_s w }{r_{\mathrm{s},\mathrm{u}}^{\beta}\gamma_{\mathrm{a,I}}} \theta_{F_{\mathrm{s},\mathrm{u}}}\right)^{-k_{F_{\mathrm{s},\mathrm{u}}}}.
\end{equation}

{\bf Remark 5}~(Access Link Rate Coverage of an SBS in OBFD Mode){\bf.}
Similar to the IBFD mode in (\ref{eq:Lap_Isu}), the Laplace transform of  {$\hat{I}_{\mathrm{s,u}}$} can be obtained as $
\mathcal{L}_{\hat{I}_{\mathrm{s},\mathrm{u}}}\left( t \right) =
\mathbb{E}_{\bar{\Phi}_{\mathrm{sO}}} 
\left[
\prod_{x \in \bar{\Phi}_{\mathrm{sO}} \setminus s} 
(1+t P_s  \;r_{\mathrm{x},\mathrm{u}}^{-\beta}
\;\theta_{F_{\mathrm{x},\mathrm{u}}})^{-k_{F_{\mathrm{x},\mathrm{u}}} }
\right]$. A closed-form expression can be given by replacing $\bar{\lambda}_{\mathrm{sI}}$ in \eqref{eq:Lap_Isu} with $\bar{\lambda}_{\mathrm{sO}}$.
Consequently, the rate coverage  at the access link of an SBS in OBFD mode can be derived  by replacing $\gamma_{\mathrm{a,I}}$ with $\gamma_{\mathrm{a,O}}$ and $F_{{I_{\mathrm{agg}}}} $ with $F_{{\hat{I}_{\mathrm{s,u}}}}$ in  {\bf Lemma~1} mentioned above. 

\subsection{Rate Coverage Analysis: Backhaul Links}

Now we derive the Laplace transforms of the interferences in the backhaul link of a given SBS operating in IBFD mode, i.e., $I_{\mathrm{s,s}}$ and $I_{\mathrm{c,s}}$. The Laplace transform of $I_{{s,s}}$ can be obtained as $\mathcal{L}_{I_{\mathrm{s},\mathrm{s}}}\left( t \right) =
\mathbb{E}_{\bar{\Phi}_{\mathrm{sI}}} 
\left[
\prod_{x \in \bar{\Phi}_{\mathrm{sI}}} 
(1+t P_s  
\;r_{\mathrm{x},\mathrm{s}}^{-\beta}
\;\theta_{F_{\mathrm{x},\mathrm{s}}})^{-k_{F_{\mathrm{x},\mathrm{s}}} }
\right]$. A closed-form expression can then be given simplifying the result in Eq.~\eqref{eq:Lap_Isu}(d) for $r_{\mathrm{s,u}} \rightarrow 0$ as follows:
\begin{align}\label{eq:Lap_Iss}
\mathcal{L}_{I_{\mathrm{s,s}}}\left( t \right)  
{=}& 
\lim_{r_{\mathrm{s,u}} \to 0} e^{ -\pi \bar{\lambda}_{\mathrm{sI}} r^2_{\mathrm{s},\mathrm{u}} 
\:_2F_1\left[k_{F_\mathrm{x,u}},-\frac{2}{\beta},
 1-\frac{2}{\beta},  -\frac{t  P_s  
\;\theta_{F_{\mathrm{x},\mathrm{u}}}}{ r_{\mathrm{s},\mathrm{u}}^{\beta}}\right] 
}, 
\nonumber\\
\stackrel{(a)}{=}&
\lim_{r_{\mathrm{s,u}} \to 0} e^{ -\pi \bar{\lambda}_{\mathrm{sI}} 
\frac{\:_2F_1
\left[1-\frac{2}{\beta}-k_{F_\mathrm{x,u}},-\frac{2}{\beta},
 1-\frac{2}{\beta}, \frac{t  P_s  
\;\theta_{F_{\mathrm{x},\mathrm{u}}}}{ r_{\mathrm{s},\mathrm{u}}^{\beta} + t  P_s  
\;\theta_{F_{\mathrm{x},\mathrm{u}}} }
\right]}{\left(r_{\mathrm{s},\mathrm{u}}^{\beta}+
t  P_s \;\theta_{F_{\mathrm{x},\mathrm{u}}}\right)^{-\frac{2}{\beta}}}
},
\nonumber\\
{=}&
\exp \left[ -\pi \bar{\lambda}_{\mathrm{sI}} 
\frac{\:_2F_1
\left[1-\frac{2}{\beta}-k_{F_\mathrm{x,u}},-\frac{2}{\beta},
 1-\frac{2}{\beta}, 1
\right]}{\left(t  P_s \;\theta_{F_{\mathrm{x},\mathrm{u}}}\right)^{-\frac{2}{\beta}}}
\right],
\nonumber\\
\stackrel{(b)}{=}&
\exp \left[-\pi \bar{\lambda}_{\mathrm{sI}} 
\left(
\frac{2 \mathcal{B} \left[k_{F_\mathrm{x,s}}+\frac{2}{\beta}, -\frac{2}{\beta}\right] }
{\beta (t {P}_s \;\theta_{F_{\mathrm{x},\mathrm{s}}})^{-\frac{2}{\beta}} }
\right)
\right],
\end{align}
where (a) follows from Pfaff identity, i.e., $$\:_2F_1(a,b,c,z)=(1-z)^{-b}\:_2F_1\left[c-a,b,c,\frac{z}{z-1}\right],$$ and (b) follows from the definition of Gauss's Hypergeometric function, i.e.,
$$\:_2F_1(a,b,c,z)=\frac{\Gamma(c)\Gamma(c-a-b)}{\Gamma(c-a)\Gamma(c-b)},$$ and applying the definition of Beta function.

The Laplace transform of $I_{\mathrm{c,s}}$ can be derived by following the steps in \eqref{eq:Lap_Icu} as follows:
\begin{align}\label{eq:Lap_Ics}
&
\mathcal{L}_{I_{\mathrm{c},\mathrm{s}}}\left( t \right) = 
\exp 
\left[ - 2\pi {\lambda}_c \int_{r_{\mathrm{c,s}}}^\infty 
\left(1 - \exp \left[ -t {P}_c  r_{\mathrm{y},\mathrm{s}}^{-\beta} \right] \right) r_{\mathrm{y},\mathrm{s}} \text{d}r_{\mathrm{y},\mathrm{s}}\right],
\nonumber\\
  &
{= \exp {\left[ - \pi {\lambda}_c \left( 
\frac{ \Gamma\left(1-\frac{2}{\beta}\right) + \frac{2}{\beta} \Gamma_u\left(-\frac{2}{\beta}, \frac{t {P}_c}  {r_{\mathrm{c,s}}^{\beta}}\right) }{(t {P}_c)^{-\frac{2}{\beta}}}
- r_{\mathrm{c,s}}^2 \right) \right]}.}
\end{align}

We apply Gil-Pelaez inversion theorem to evaluate the rate coverage probability for an SBS in the backhaul link as shown below.
\begin{lemma}[Backhaul Link Rate Coverage of an SBS in IBFD Mode]
With Gamma-distributed interfering links, the coverage probability of the backhaul link of an SBS in IBFD mode can be derived  as follows:
\begin{align}\label{cov_bl}
&
{
\mathbb{P}(\mathrm{SIR}_{\mathrm{b,I}}>{\gamma}_{\mathrm{b,I}})}
\nonumber\\&
{
=
2 \pi  \lambda_c
\int_0^\infty F_{{I_{\mathrm{agg}}}} \left( 
\frac{{P}_c \; \; }{{\gamma}_{\mathrm{b,I}} r_{\mathrm{c,s}}^{\beta}}  
-I_{\mathrm{SI}}
\right)
 r_{\mathrm{c,s}} e^{-\pi  \lambda_c {r_{\mathrm{c,s}}}^2}
\text{d}r_{\mathrm{c,s}},}
\nonumber
\end{align}
where  $F_{I_{\mathrm{agg}}} (\cdot)$ can be derived  as
\begin{align}
&F_{{I_{\mathrm{agg}}}} \left( 
\frac{{P}_c \;\; }{{\gamma}_{\mathrm{b,I}} r_{\mathrm{c,s}}^{\beta}} 
-I_{\mathrm{SI}}
\right)
\nonumber\\&
=
\frac{1}{2} -  \frac{1}{\pi}\int_0^\infty \mathrm{Im} \left[ \frac{\mathcal{L}_{I_{\mathrm{c,s}}}\left( -jw \right)
\mathcal{L}_{I_{\mathrm{s,s}}}\left( -j w \right)}
{ e^{j w \left( 
\frac{{P}_c \;\; }{{\gamma}_{\mathrm{b,I}} r_{\mathrm{c,s}}^{\beta}} 
-I_{\mathrm{SI}}
\right)}}
 \right] \frac{\text{d} w}{w}.
\end{align}
\end{lemma}

{\bf Remark 6}~(Backhaul Link Rate Coverage for an SBS in OBFD Mode){\bf.}
Similar to the IBFD mode, the Laplace transform of $I_{\mathrm{s,s}}$ and $I_{\mathrm{c,s}}$ in OBFD mode can be given as in (\ref{eq:Lap_Iss}) and \eqref{eq:Lap_Ics}, respectively. The received SIR at the backhaul link of an SBS operating in OBFD mode (i.e., $\mathrm{SIR}_{\mathrm{b,O}}$) is thus similar to that of an SBS operating in the IBFD mode in \eqref{eq:SINRbI}. Hence,  \eqref{cov_bl} can be used for the OBFD case after replacing ${\gamma}_{\mathrm{b,I}}$ with ${\gamma}_{\mathrm{b,O}}$ and substituting $I_{\mathrm{SI}}=0$.

\section{Simplified Rate Coverage Expressions: Specific Scenarios and Approximations}

In this section, we provide simplified  expressions for an interference-limited scenario with Rayleigh fading  at the access  links. Moreover, we exploit an approximation of the Gauss's hypergeometric function to further simplify the rate coverage expressions in the access link. Based on the simplified expressions and considering ideal backhaul rate coverage, we derive closed-form expressions for (i)~the  value of $q$ that maximizes the rate coverage probability of a typical user; (ii)~the  value of $q$ at which the the rate coverage probability of all SBSs operating in IBFD  and OBFD mode in a small cell network can be balanced.

\subsection{Simplified Rate Coverage for Rayleigh Fading}
The simplified rate coverage  probability expressions in the access link of an IBFD SBS can be derived in closed-form  for a typical Rayleigh fading interference-limited scenario as described in the following.
\begin{lemma}[Access Link Rate Coverage for an SBS in IBFD Mode in Rayleigh Fading Channels]
For Rayleigh fading channels, i.e., $F_{\mathrm{s,u}}  \sim \mathrm{Gamma}(1,\theta_{F_{\mathrm{s,u}}})$ and interference-limited regime, the rate coverage probability experienced at the access link of an IBFD SBS can be simplified as follows:
\begin{align}\label{derivation1}
&\mathbb{P}(\mathrm{SIR}_{\mathrm{a,I}}>\gamma_{\mathrm{a,I}})
\stackrel{(a)}{=}
\mathbb{E}\left[
\exp\left(-\frac{\gamma_{\mathrm{a,I}} r_{\mathrm{s,u}}^{\beta}}{P_s \theta_{F_{\mathrm{s,u}}} } (I_{\mathrm{c,u}} +I_{\mathrm{s,u}})\right)\right],
\nonumber\\
&
\stackrel{(b)}{=}\int_0^\infty 
\mathcal{L}_{I_{\mathrm{s,u}}}\left( 
\frac{\gamma_{\mathrm{a,I}} r_{\mathrm{s,u}}^{\beta}}{P_s \theta_{F_{\mathrm{s,u}}} }
 \right) 
\mathcal{L}_{I_{\mathrm{c,u}}}
\left( \frac{\gamma_{\mathrm{a,I}} r_{\mathrm{s,u}}^{\beta}}{P_s \theta_{F_{\mathrm{s,u}}} } \right) f_{r_{\mathrm{s,u}}}(r_{\mathrm{s,u}}) \text{d}r_{\mathrm{s,u}},
\nonumber\\
&
\stackrel{(c)}{=} 
{
\frac{\lambda_s}{   \bar{\lambda}_{\mathrm{SI}}
(\:_2F_1\left[1,-\frac{2}{\beta},
 1-\frac{2}{\beta},  -\gamma_{\mathrm{a,I}} \right]-1)
+ \lambda_s+ A}},
\end{align}
where $A={\Gamma\left(1-\frac{2}{\beta}\right) {\lambda}_c}{\left(\frac{\gamma_{\mathrm{a,I}} {P}_c }{P_s \theta_{F_{\mathrm{s,u}}} } \right)^{\frac{2}{\beta}} }$ and is independent of $q$.
\end{lemma}
Note that (a) follows from the CDF of the exponential distribution with average power $\theta_{F_{\mathrm{s,u}}}$, (b) follows from the definition of Laplace transform and the independence of $I_{\mathrm{c,u}}$ and $I_{\mathrm{s,u}}$, and (c) follows from substituting \eqref{eq:Lap_Isu} and \eqref{eq:Lap_Icu} and evaluating the integral.

The backhaul rate coverage expressions can also be simplified for Rayleigh fading channels due to the simplification of  Beta function in $\mathcal{L}_{I_{\mathrm{s,s}}}(t)$ as follows:
\begin{align}\label{eq:Lap_Iss1}
&\mathcal{L}_{I_{\mathrm{s,s}}}\left( t \right)  
{=} 
\exp \left[- 
\frac{2 \pi^2 \bar{\lambda}_{\mathrm{SI}}  \mathrm{csc}(\frac{2 \pi}{\beta}) }
{(t {P}_s \;\theta_{F_{\mathrm{x},\mathrm{s}}})^{-\frac{2}{\beta}} \beta}
\right].
\end{align}

\subsection{Approximate Rate Coverage for Gamma/Rayleigh Fading}

From  \cite[13,  Ch.  5,  Eq.  (2)]{erdeli},  we  have
$\:_2F_1\left[-\frac{2}{\beta}, k,
 1-\frac{2}{\beta},  j x \right]
\approx 1-\frac{j 2 x k}{\beta -2}$. With this approximation, the rate coverage  expressions in the access link can be further simplified for both Gamma and Rayleigh fading channels. The Rayleigh fading case is detailed below.

\begin{corollary}[Approximate Access Link Rate Coverage for an SBS in IBFD/OBFD Mode Under Rayleigh Fading Channels]
With the aforementioned approximation, the rate coverage probability experienced at the access link of an SBS in IBFD mode can be simplified as follows:
\begin{align}\label{derivation11}
&\mathbb{P}(\mathrm{SIR}_{\mathrm{a,I}}>\gamma_{\mathrm{a,I}})
{\approx} 
{
\frac{\lambda_s}{   \bar{\lambda}_{\mathrm{SI}} 
\frac{2}{\beta-2} \gamma_{\mathrm{a,I}}
+ \lambda_s+
{\Gamma\left(1-\frac{2}{\beta}\right) {\lambda}_c}{(\frac{\gamma_{\mathrm{a,I}} {P}_c }{P_s \theta_{F_{\mathrm{s,u}}} } )^{\frac{2}{\beta}} }}}.
\end{align}
Similarly, the approximate access link rate coverage expression  for OBFD mode  can be derived as follows:
\begin{align}\label{derivation11}
\mathbb{P}(\mathrm{SIR}_{\mathrm{a,O}}>\gamma_{\mathrm{a,O}})
{\approx} 
{
\frac{\lambda_s}{   \bar{\lambda}_{\mathrm{SO}} 
\frac{2}{\beta-2} \gamma_{\mathrm{a,O}}
+ \lambda_s}}.
\end{align}
\end{corollary}
In the interference-limited regime and perfect backhaul coverage, it can be observed that the rate coverage in the access link of an SBS in IBFD mode is vulnerable to both the transmit powers and intensities of CNs and  SBSs. On the other hand, the rate coverage of an SBS in OBFD mode is  independent of the transmit powers of CNs or SBSs. Furthermore, if the average load per CN is less than $\mathcal{S}$, the rate coverage of an SBS in OBFD mode becomes independent of the  intensities of CNs or SBSs. This can be verified by substituting the value of $ \bar{\lambda}_{\mathrm{SO}}$ in {\bf Corollary~1}. 
These observations become more evident  when $M\rightarrow \infty$ and in turn $\gamma_{\mathrm{b,I}}, \gamma_{\mathrm{b,O}} \rightarrow 0$.

\subsection{Perfect Backhaul Coverage: Large Values of $M$}
The ideal backhaul rate coverage can be realized when the ratio $\frac{M}{\mathrm{min}(\mathcal{N}_s, \mathcal{S})}$ becomes very large,  i.e., $\gamma_{\mathrm{b,I}}$ and $\gamma_{\mathrm{b,O}}$ become very small as can be observed from \eqref{gammabI}.
In such a case, we can customize the value  of $q$ in closed-form such that either  the user's rate coverage is maximized (i.e., $q^{*}$) or the rate coverage of all SBSs in the network (in IBFD or OBFD mode) becomes balanced (i.e., $q^{\mathrm{balance}}$).

{The optimization of rate coverage leads to optimized network performance when each SBS can adaptively serve its associated user in both the IBFD and OBFD modes. That is, each SBS can set the optimal proportion $q^{*}$ to switch between IBFD and OBFD modes. Nonetheless,  all SBSs may not have the capability to adaptively switch into different modes or some BSs can only operate in OBFD mode (could be due to unavailability of advanced SI cancellation circuitry or any other implementation issues). In such a case, a user connected to an IBFD SBS will experience  significantly different rate coverage probability from a user connected to an OBFD SBS depending on the proportion of IBFD and OBFD SBSs in the network. Consequently,  it will be important  to set the proportion of IBFD SBSs in the network such that they do not hurt the rate coverage probability of users associated to OBFD SBSs. The proportion which balances the rate coverage in all IBFD and OBFD SBSs $q^{\mathrm{balance}}$ can be decided in advance so that each user can experience the same rate coverage.}

In the following,  we derive both the $q^{*}$ and  $q^{\mathrm{balance}}$ in closed-form.

\begin{corollary}[Balanced Rate Coverage  with Perfect Backhaul Rate Coverage]
In order to balance the rate coverage among all IBFD and OBFD SBSs in the network, we equate  $\mathbb{P}(\mathrm{SIR}_{\mathrm{a,O}}>\gamma_{\mathrm{a,O}})$ and $\mathbb{P}(\mathrm{SIR}_{\mathrm{a,I}}>\gamma_{\mathrm{a,I}})$ that leads to
$
 \bar{\lambda}_{\mathrm{SO}} 
\:_2F_1\left[1,-\frac{2}{\beta},
 1-\frac{2}{\beta},  -\gamma_{\mathrm{a,O}} \right]
-\bar{\lambda}_{\mathrm{SI}} 
\:_2F_1\left[1,-\frac{2}{\beta},
 1-\frac{2}{\beta},  -\gamma_{\mathrm{a,I}} \right]
=
A+ \bar{\lambda}_{\mathrm{SO}} -\bar{\lambda}_{\mathrm{SI}}.
$
Substituting the expressions of $\bar{\lambda}_{\mathrm{SI}} $ and $\bar{\lambda}_{\mathrm{SO}} $ from {\bf Remark~2} and {\bf Remark~3} we can derive $q$ as follows:
\begin{equation}
\Scale[1]
{
q^{\mathrm{balance}} 
=
\frac{_2F_1\left[1,-\frac{2}{\beta},
 1-\frac{2}{\beta},  -\gamma_{\mathrm{a,O}} \right]-1-\frac{A}{\lambda_c \mathrm{min}(\lambda_s/\lambda_c, \mathcal{S})  }}{\:_2F_1\left[1,-\frac{2}{\beta},
 1-\frac{2}{\beta},  -\gamma_{\mathrm{a,O}} \right]+\:_2F_1\left[1,-\frac{2}{\beta},
 1-\frac{2}{\beta},  -\gamma_{\mathrm{a,I}} \right]-2}.}
\end{equation}
\end{corollary}
Also, in this case, the access link rate coverage of a typical user defined as $\mathcal{C}_u=q\mathbb{P}(\mathrm{SIR}_{\mathrm{a,I}}>\gamma_{\mathrm{a,I}}) +(1-q) \mathbb{P}(\mathrm{SIR}_{\mathrm{a,O}}>\gamma_{\mathrm{a,O}})$ can be optimized with respect to $q$. However, the presence of $\:_2F_1(\cdot)$ in the expression makes it difficult to obtain direct  insights. As such, we  approximate   $\:_2F_1(\cdot)$  as in  {\bf Corollary~1}. In the following, we derive a more simplified closed-form expression of $q^{\mathrm{balance}}$.

\begin{corollary}[Approximate Balanced Rate Coverage with Perfect Backhaul Rate Coverage]
In order to balance the rate coverage in IBFD as well as OBFD mode, we equate $\mathbb{P}(\mathrm{SIR}_{\mathrm{a,O}}>\gamma_{\mathrm{a,O}})$ and $\mathbb{P}(\mathrm{SIR}_{\mathrm{a,I}}>\gamma_{\mathrm{a,I}})$.

Substituting the expressions of $\bar{\lambda}_{\mathrm{SI}} $ and $\bar{\lambda}_{\mathrm{SO}} $ from {\bf Remark~2} and {\bf Remark~3} we can derive $q$ as follows:
\begin{equation}
q^{\mathrm{balance}} 
=
\frac{\gamma_{\mathrm{a,O}}-\frac{A(\beta-2)}{2\lambda_c \mathrm{min}(\lambda_s/\lambda_c, \mathcal{S}) }}{ \gamma_{\mathrm{a,I}} + 
\gamma_{\mathrm{a,O}}}.
\end{equation}
\end{corollary}
Note that $q^{\mathrm{balance}} $ is inversely proportional to $A$ and in turn the transmission powers of CNs and SBSs, i.e., $\frac{P_c}{P_s}$. Moreover,  when the average load per CN is less than the $\mathcal{S}$, $q^{\mathrm{balance}}$ becomes inversely proportional to the factor $\frac{\lambda_s}{\lambda_c}$. That is, increasing the intensity/power of SBSs requires an increase in the fraction of SBSs operating in IBFD mode to maintain a fair rate coverage at all SBSs in a small cell network.

Similarly, the rate coverage of a typical user in the access link can be optimized with respect to $q$ as detailed below.
\begin{corollary}[Optimal Rate Coverage of a Typical User with Perfect Backhaul Rate Coverage]
{
Assuming perfect backhaul coverage, using the relation $\mathcal{C}_u=q \mathcal{C}_{\mathrm{I}} +(1-q) \mathcal{C}_{\mathrm{O}}$ for rate coverage
and substituting $\mathcal{C}_{\mathrm{I}}$ and $\mathcal{C}_{\mathrm{O}}$ from eq. (31) and eq. (32) yields
\begin{equation}
\mathcal{C}_u=  {
\frac{q\lambda_s}{   \bar{\lambda}_{\mathrm{SI}} 
\frac{2}{\beta-2} \gamma_{\mathrm{a,I}}
+ \lambda_s+A
}}
+
{
\frac{ (1-q)\lambda_s}{   \bar{\lambda}_{\mathrm{SO}} 
\frac{2}{\beta-2} \gamma_{\mathrm{a,O}}
+ \lambda_s}}
\end{equation}
Since  $\bar{\lambda}_{\mathrm{SI}}=q \times \mathrm{min}(\lambda_s/\lambda_c, \mathcal{S}) \lambda_c$ and $\bar{\lambda}_{\mathrm{SO}}= (1-q) \times \mathrm{min}(\lambda_s/\lambda_c, \mathcal{S}) \lambda_c$, we can write
\begin{equation}
\mathcal{C}_u=  {
\frac{q\lambda_s}{   q 
c \gamma_{\mathrm{a,I}}
+ \lambda_s+
A}}
+
{
\frac{ (1-q)\lambda_s}{   (1-q)  
c \gamma_{\mathrm{a,O}}
+ \lambda_s}},
\end{equation}
}
where $c=\frac{2 \lambda_c \times \mathrm{min}(\lambda_s/\lambda_c, \mathcal{S})}{\beta-2}$.
Now  differentiating  each term with respect to $q$   yields
\begin{equation}
\frac{d \mathcal{C}_u}{d q}=
\lambda_s 
\left(\frac{A + \lambda_s}{(A + c q \gamma_{\mathrm{a,I}} + \lambda_s)^2} 
 - \frac{\lambda_s}{(c \gamma_{\mathrm{a,O}} (1 - q ) + \lambda_s)^2}\right).
\end{equation}
 An optimum $q$ for a typical user can then be given as follows:
\begin{equation}\label{corollary5}
\Scale[1]
{q^*=
\frac{(A + \lambda_s)(c^2 \gamma_{\mathrm{a,O}}^2 +d) \pm (c(A + c \gamma_{\mathrm{a,I}})  \gamma_{\mathrm{a,O}}+d) \sqrt{\lambda_s(A + \lambda_s)}}
{c^2(-\gamma_{\mathrm{a,I}}^2 \lambda_s + \gamma_{\mathrm{a,O}}^2(A+\lambda_s) )},}
\end{equation}
where $d=c (\gamma_{\mathrm{a,I}} +\gamma_{\mathrm{a,O}}) \lambda_s$.
Note that $(A + \lambda_s)(c^2 \gamma_{\mathrm{a,O}}^2 +d) > {c^2(-\gamma_{\mathrm{a,I}}^2 \lambda_s + \gamma_{\mathrm{a,O}}^2(A+\lambda_s) )}$ and $0 \leq q \leq 1$, thus the root with a negative sign is feasible only.
\end{corollary}
By replacing $c={\lambda_c \times \mathrm{min}(\lambda_s/\lambda_c, \mathcal{S})}$, $\gamma_{\mathrm{a,I}}=\:_2F_1[1,-\frac{2}{\beta},
 1-\frac{2}{\beta},  -\gamma_{\mathrm{a,I}} ]-1$, and $\gamma_{\mathrm{a,O}}=\:_2F_1[1,-\frac{2}{\beta},
 1-\frac{2}{\beta},  -\gamma_{\mathrm{a,O}} ]-1$, \eqref{corollary5} can be used to obtain $q^*$ without approximation.

\section{Backhaul Interference Mitigation Techniques}

Since the backhaul interference incurred at a user terminal appears to be a fundamental bottleneck in  the performance of in-band wireless backhauling, some kind of coordination between CNs and SBSs is crucial to achieve the performance gains of IBFD over OBFD mode. 
In this context, this section first discusses a distributed backhaul aware mode selection mechanism which provides some further insights into selecting the proportion of in-band and out-of-band SBSs ($q$) as a function of network parameters. We then discuss two   solution techniques that can potentially enhance the performance of in-band wireless backhauling, namely, Backhaul Interference-Aware (BIA) power control at CNs and Interference Rejection (IR) from the CN the serving SBS of the typical user is associated with. Since  the locations of the BSs are relatively fixed, the aforementioned backhaul interference management  techniques 
can be easily implemented in practice.

\subsection{Distributed Backhaul-Aware IBFD/OBFD Mode Selection}

Based on the received power at a certain user from both the nearest CN and its serving SBS, the serving SBS independently chooses its mode, i.e., IBFD or OBFD mode. In practice, this mode selection can also be performed by the user as the interference estimation could be much easier at the user's end in the downlink. The user can then inform its designated SBS about the feasible mode of operation. 

Specifically, the SBS chooses the IBFD mode if the received signal power from that SBS at the user is sufficiently higher than the power received from the strongest interfering CN. In such a case, using more radio resources by switching to the OBFD mode is not necessary. On the other hand, if the signal power at the user from the strongest interfering CN is comparable or higher than the useful signal power received from its serving SBS, the OBFD mode of operation is selected. The  mode selection criterion can be written mathematically as:
\begin{align}
\begin{cases}
\frac{P_s r_{\mathrm{s0,u}}^{-\beta}}{{P}_c r_{\mathrm{c0,u}}^{-\beta}} \geq \tau, & \mathrm{IBFD~mode}\\
\frac{P_s r_{\mathrm{s0,u}}^{-\beta}}{{P}_c r_{\mathrm{c0,u}}^{-\beta}} < \tau, & \mathrm{OBFD~mode}
\end{cases}
\end{align}
where $\tau$ is a threshold that can be different for each user and can  be chosen as a function of the target rate requirement of that user, $r_{\mathrm{s0,u}}$ is the distance from the user to its serving SBS and $r_{\mathrm{c0,u}}$ is the distance from the same user to its closest CN. 
Given the mode selection criterion, the expression for $q$ can be derived for path-loss only environment as follows:
\begin{align}\label{q}
q &
= \mathbb{E}_{r_{\mathrm{s0,u}}} \left[ \mathbb{P}\left[ r_{\mathrm{c0,u}} > \left(\frac{\tau {P}_c}{P_s}\right)^{\frac{1}{\beta}} r_{\mathrm{s0,u}} \right] \right],\nonumber\\
                                &\stackrel{(a)}{=} \int_0^\infty \exp \left[ -\pi \lambda_c \left(\frac{\tau {P}_c}{P_s}\right)^{\frac{2}{\beta}} r^2 \right] \cdot 2\pi \lambda_s r \exp \left[ -\pi \lambda_s r^2 \right] \text{d}r,\nonumber\\
                                &= \left({1+\frac{\lambda_c}{\lambda_s} \left(\tau   \frac{ {P}_c}{P_s}\right)^{\frac{2}{\beta}}}\right)^{-1},
\end{align}
where (a) follows by using the distribution of distance to the nearest point in a PPP.
Note that, $\tau$ is a design parameter which plays a key role in controlling the trade-off between the rate coverage and the backhaul interference. Intuitively, in case when the backhaul interference from the nearest CN is more dominant,  $\tau$ can be comparable to
$\gamma_{\mathrm{a,I}}$ or $\gamma_{\mathrm{a,O}}$. However, in the other case $\tau$  should be selected sufficiently higher than 
$\gamma_{\mathrm{a,I}}$ or $\gamma_{\mathrm{a,O}}$ to consider  the impact of additional co-tier and cross-tier interferences.

{As has been mentioned in previous sections, for a given set of network parameters, a network designer can determine the optimized value of $q$ given the statistical knowledge/distributions of the intensity and locations of the CNs, SBSs, users, and channel gains among them. The optimal value of $q$ can then be broadcast to all SBSs in the system. However,  a given SBS can also decide the mode in a distributed manner (e.g., by using \eqref{q}).
Nonetheless, since the decisions of a SBS are dependent only on the statistics of its users' channels and the incurred backhaul interference, the distributed mode selection can lead to performance degradation (especially if the  threshold $\tau$ is not  optimized). This  will be demonstrated via numerical results in Section VII.}

\subsection{Interference Rejection}

For massive MIMO systems, backhaul interference at a given user from the  CN its SBS is associated with can be completely rejected. This can be done by designing precoders  assuming full knowledge of the channel state information (CSI) between that CN and all users within its coverage~\cite{ibfd3}.
For example, if $\mathbf{H}_{\mathrm{c,u}}$ is the $1 \times M$ channel vector between the CN and a given user $u$, the interference at $u$ from that CN can be canceled (or rejected) by multiplying the transmit signal (i.e., beamforming) from the CN by the matrix $\mathbf{R}(\mathbf{H}_{\mathrm{c,s}}\cdot\mathbf{R})^{\dagger}$ where $\mathbf{H}_{\mathrm{c,s}}$ is the channel vector between the serving SBS of $u$ and its serving CN, $(\cdot)^{\dagger}$ denotes the pseudo inverse, and $\mathbf{R}$ is calculated to satisfy the condition
$
\mathbf{H}_{\mathrm{c,u}} \cdot\mathbf{R} = \mathbf{0}.
$
{ 
In this case, the channel gain between this CN and the user served by its associated SBS is $\mathbf{H}_{\mathrm{c,u}} \mathbf{R}(\mathbf{H}_{\mathrm{c,s}}\cdot\mathbf{R})^{\dagger} = \mathbf{0} (\mathbf{H}_{\mathrm{c,s}}\cdot\mathbf{R})^{\dagger} = \mathbf{0}$.
Note that, the number of antenna $M$ needs to be greater than or equal to $2 \times \min(\mathcal{N}_s, \mathcal{S})$ (i.e., the total number of SBS and the associated users under the assumption that  one user is served per SBS at a time). 
}
As such, the interference in \eqref{Icu} reduces to the following:
\begin{equation}
I_{\mathrm{c},\mathrm{u}}=\sum_{y\in {\Phi}_{c} } {P_c r_{\mathrm{y},\mathrm{u}}^{-\beta} }  = \sum_{y\in {\Phi}_{c} \setminus c} {P}_c \; r_{\mathrm{y},\mathrm{u}}^{-\beta}.
\end{equation}
Hence, using (\ref{eq:Lap_Ics}), the Laplace transform of the backhaul interference can be given as follows:
\begin{equation}
\mathcal{L}_{I_{\mathrm{c},\mathrm{u}}}\left( t \right) = \mathcal{L}_{I_{\mathrm{c},\mathrm{s}}}\left( t \right) |_{r_{\mathrm{c},\mathrm{s}} = r_{\mathrm{c},\mathrm{u}}}.
\end{equation}
 
\subsection{Backhaul Interference-Aware (BIA) Power Control at CNs}
BIA power control scheme requires the following information  at a given CN $c$, i.e.,
\begin{itemize}
\item the cumulative interference at a given IBFD SBS $s$ associated with that CN $c$ ($I_{\mathrm{agg}}=I_{\mathrm{c,s}}+I_{\mathrm{s,s}}+I_{\mathrm{SI}}$), 
\item the desired link between CN $c$ and the IBFD SBS $s$, and
\item the desired target SIR of the backhaul link $\gamma_{\mathrm{b,I}}$.
\end{itemize}
Based on these informations, a given CN reduces its power to a given IBFD SBS from $\frac{P_c}{\mathrm{min}(\mathcal{N}_s,\mathcal{S})}$ to a level that can strictly ensure $\gamma_{\mathrm{b,I}}$. Otherwise, the CN continues to transmit with power $\frac{P_c}{\mathrm{min}(\mathcal{N}_s,\mathcal{S})}$. 
This power reduction can potentially reduce the cross-tier interference experienced at a user associated with IBFD SBS $s$.
Mathematically, the power level of a CN can then be defined as follows:
\begin{equation}
P^{s}_{\mathrm{cn}}=\mathrm{min}\left(\frac{P_c}{\mathrm{min}(\mathcal{N}_s,\mathcal{S})},\frac{ \gamma_{\mathrm{b,I}} I_{\mathrm{agg}}}{r_{\mathrm{c,s}}^{-\beta} S_{\mathrm{c,s}}}\right).
\end{equation}
However, since the factor $\mathrm{min}(\mathcal{N}_s,\mathcal{S})$ is already considered in the definition of $\gamma_{\mathrm{b,I}}$ (see \eqref{backhaul} and \eqref{gammabI}), thus we can write:
\begin{equation}
P^{s}_{\mathrm{cn}}=\mathrm{min}\left({P_c},\frac{ \gamma_{\mathrm{b,I}} I_{\mathrm{agg}}}{r_{\mathrm{c,s}}^{-\beta} S_{\mathrm{c,s}}}\right).
\end{equation}

In practice, the aforementioned CSI is most likely  available at all SBSs and thus the calculation of the desired power from CN, i.e.,  $\frac{ \gamma_{\mathrm{b,I}} I_{\mathrm{agg}}}{r_{\mathrm{c,s}}^{-\beta} S_{\mathrm{c,s}}}$ can be  performed by SBS and then forwarded to CN. Based on the load of associated SBSs per CN, the CN can  decide the transmit power level.
Note that the gains of BIA power control scheme are expected to be more evident in the following scenarios:
\begin{itemize}
\item reduced number of SBSs per CN,
\item low values of $\gamma_{\mathrm{b,I}}$, and
\item high total transmit power $P_c$ of a given CN.
\end{itemize}
In the above scenarios,  equal transmit power allocation per backhaul stream at CN results in unnecessarily high transmission power. As such, the significance of applying the adaptive power control at a CN per IBFD stream is evident.

\begin{figure*}
\captionsetup[figure]{labelformat=empty}
\centering
\begin{minipage}{.5\textwidth}
  \centering
  \includegraphics[width = 3.3in]{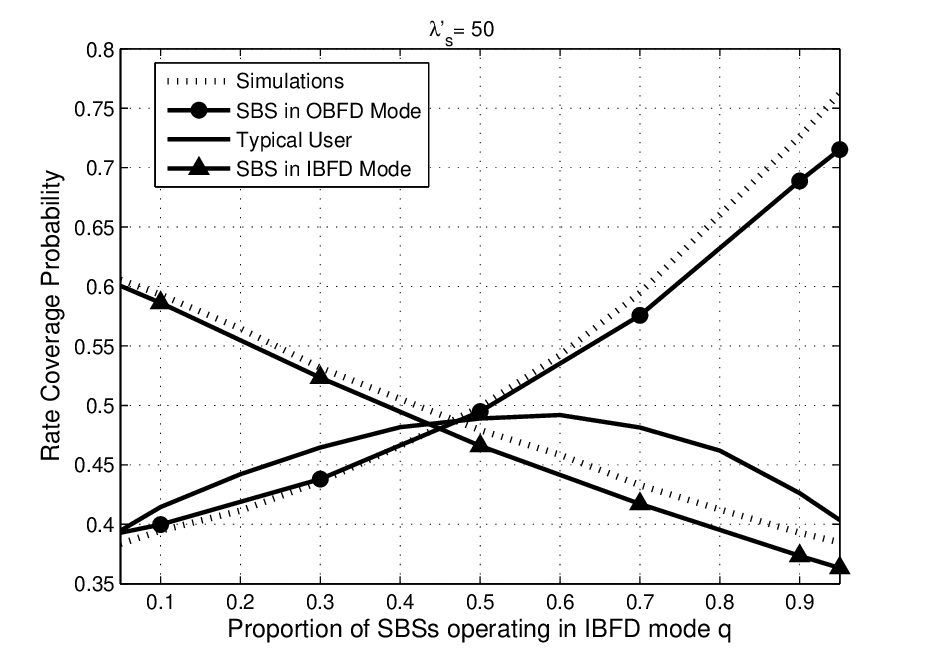}
\caption*{(a)}
\end{minipage}%
\begin{minipage}{.5\textwidth}
  \centering
  \includegraphics[width = 3.3in]{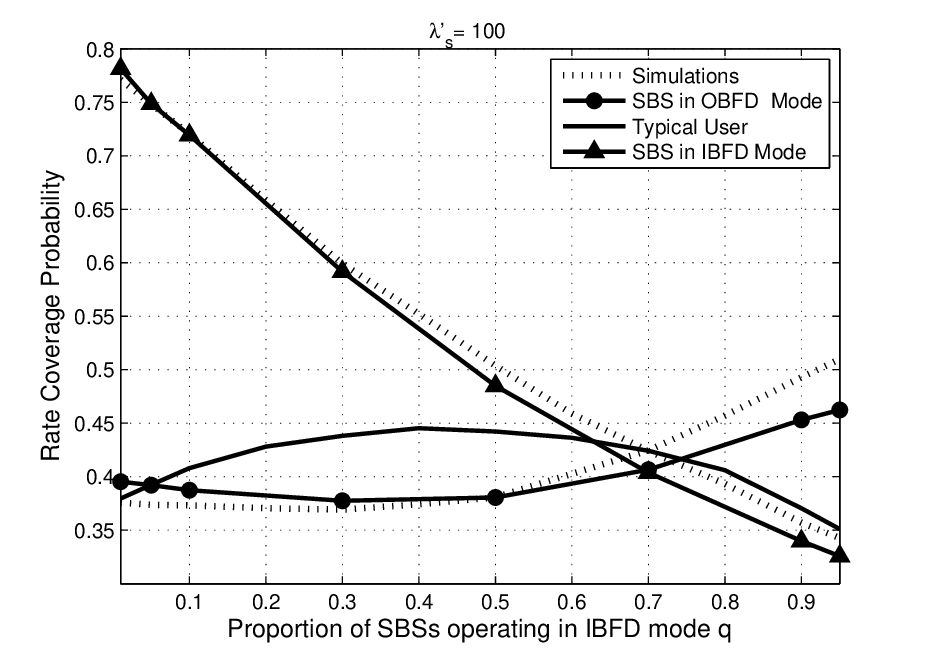}
\caption*{(b)}
\end{minipage}
\captionsetup[figure]{labelformat=default}
\caption{Rate coverage probability of SBSs operating in IBFD mode and OBFD mode as well as a typical user in the downlink as a function of the proportion of IBFD SBSs considering two different intensities of SBSs (for $\lambda_c'=10$): (a)~$\lambda^\prime_s=50$ and (b)~$\lambda^\prime_s=100$.}
\label{fig1}
\end{figure*}

\section{Numerical and Simulation Results}

In this section, we validate the accuracy of the derived expressions. The downlink coverage probability of a typical user and an SBS in both IBFD and OBFD modes  is investigated in terms of  the density of SBSs, the SI value of the IBFD SBSs, and the transmit power of the CN.  Performance trade-offs are characterized and insights are extracted related to the feasibility and selection of in-band or out-of-band backhaul modes for FD small cells. 

For our simulation results, we consider the path-loss exponent $\beta=4$ and the intensity of CNs as $\lambda_c'=10$. Note that the simulations consider the shadowing phenomenon explicitly. As such, the used intensities for CNs and SBSs are the intensities (i.e., $\lambda^\prime_c$ and $\lambda^\prime_s$) defined prior to using the displacement theorem in {\bf Remark~1}. The total transmit power available at each CN and SBS is taken as $P_c$ = 10W and $P_s$ = 2W, respectively. The number of available antennas at each CN is $M=500$ and the number of supported backhaul streams per CN is $S=50$. The  total desired rate is set as $R_{\mathrm{th}}=1$~bps/Hz. The SI cancellation value ${\xi}$ is set to 120~dB. Log-Normal shadowing channel for the  desired and interfering access links  are modeled with parameters $\mu=1$ and $\sigma=2$. The Gamma-distributed fading channel powers experienced at the SBS and at the user are  modeled with shape parameters $k=0.5$ and $k=2$, respectively, with the average channel fading power $k\theta=1$. 
The values of the above-mentioned parameters remain the same unless stated otherwise.

{In the Monte-Carlo simulations, the SBSs, CNs, and users are generated randomly in a circular cellular region in each iteration. Their numbers follow a Poisson distribution while their locations are randomly  distributed in the cellular region following a uniform distribution. In each iteration, based on the randomly generated channel gains (composed of distance-based path-loss, shadowing, and fading) of access and backhaul links, different kind of interferences and in turn the attained SINR at the access and backhaul links is calculated. Since we assume perfect channel estimation and zero-forcing beamforming,  the impact of pilot-contamination and intra-cell interference is ignored in the simulations. Moreover, in case of interference rejection, we ignore the backhaul interference from the  CN the SBS of a typical user is associated with.}

\subsection{Rate Coverage vs. Fraction of  SBSs in IBFD/OBFD Mode}
\figref{fig1} depicts the rate coverage probability of both the SBSs operating in the IBFD mode and that are operating in the OBFD mode as a function of their corresponding fractions (i.e., $q \lambda^\prime_s$ and $(1-q)\lambda^\prime_s$, respectively). Simulation results validate the accuracy of the derived expressions and demonstrate that the impact of the considered mean load approximation (i.e., $\mathbb{E}[\mathcal{N}_s]=\lambda^\prime_s/\lambda_c'$) on the gap between the simulation and analytical results is negligible. 
Note that the rate coverage of both kinds of SBSs (i.e., operating in IBFD and OBFD modes) tends to degrade with the increase in the corresponding proportion (i.e., $q$ and $1-q$, respectively). This is due to the increased co-tier interference at their corresponding access links as the intensity of the interfering SBSs increases. This trend remains valid for any intensity of SBSs as can be seen for both $\lambda^\prime_s=50$ and $\lambda^\prime_s=100$ in \figref{fig1}(a) and (b), respectively.  

{\figref{fig1} also shows the higher rate coverage  of  SBSs in IBFD mode when the intensity of SBSs (i.e., $\lambda^\prime_s$) is high. For example, at $\lambda^\prime_s=100$, 10\% of IBFD SBSs in the network experience a much higher (70\%) rate coverage compared to the case when $\lambda^\prime_s=50$. Moreover, comparing the rate coverage of 10\% of SBSs in the IBFD mode  and 10\% of SBSs in the OBFD mode, it is observed that the IBFD SBSs outperform the OBFD SBSs.
The underlying reason is that both the access and backhaul links of 10\% IBFD SBSs do not experience any interference from the  large proportion of OBFD SBSs. However, the backhaul links of 10\% of OBFD SBSs are vulnerable to the interference caused by  90\% of   IBFD SBSs.
In particular, due to the co-tier interference from the large proportion of IBFD SBSs, the  OBFD SBSs  may not enhance their rate coverage significantly even for low proportion of OBFD SBSs in the network.}

In addition, \figref{fig1} depicts that a small cell network of 100\% IBFD SBSs or 100\%  OBFD SBSs may not be beneficial. Instead,  an appropriate value of $q$ can be selected that enhances either the overall rate coverage of a typical user or ensures a fair rate coverage at all SBSs in the network. For instance, when $\lambda^\prime_s=100$, {a proportion of 70\%-30\% IBFD-OBFD SBSs in the network achieves a balanced rate coverage.  On the other hand, when $\lambda^\prime_s=50$, a proportion of 45\%-55\% IBFD-OBFD SBSs  achieves a balanced rate coverage. 

Finally, \figref{qopt} demonstrate the accuracy of $q^*$  derived in Eq.~(38) for the case of perfect backhaul coverage.  The derived value of $q^*$ closely matches the optimal value of $q$ obtained through simulations.

\begin{figure}[h]
\centering
\includegraphics[width = 3.3in]{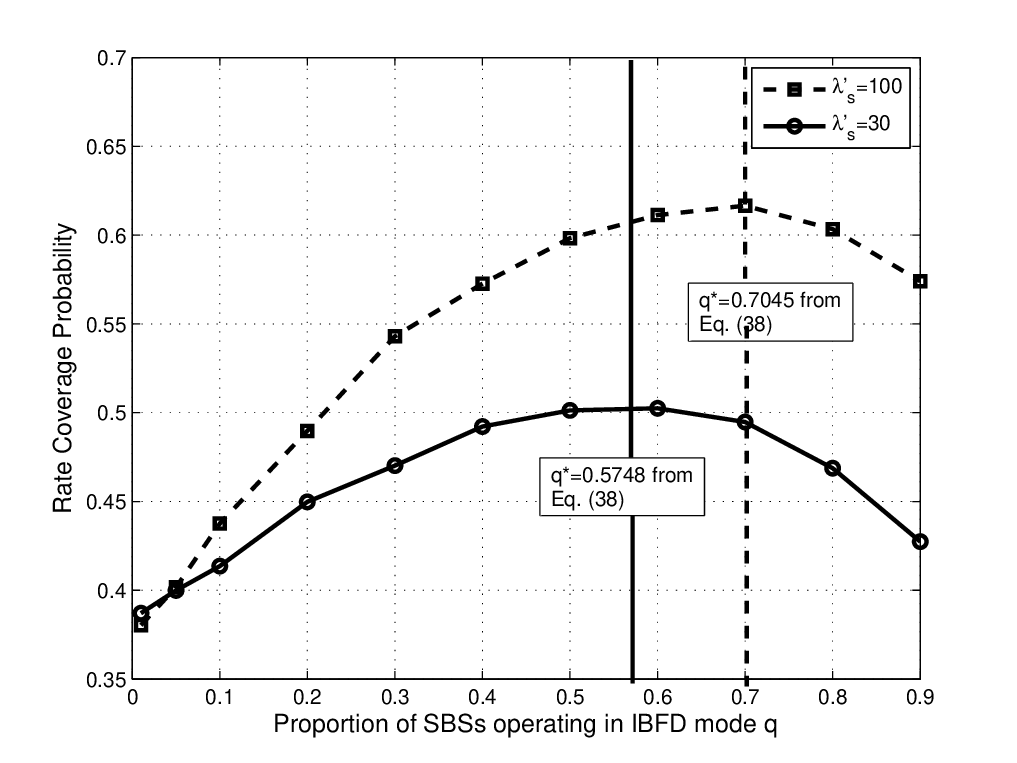}
\caption{With perfect backhaul coverage, the rate coverage probability of a typical user in the downlink as a function of $q$, $\lambda^{\prime}_c=10$.} 
\label{qopt}
\end{figure}

\subsection{Rate Coverage vs. Intensity of  SBSs}
\figref{fig2} shows the  downlink rate coverage  of  an IBFD and an OBFD SBS  as a function of the intensity of SBSs in the network. In general, it can be seen that, for a given $q$, an increasing number of SBSs  reduces the rate coverage of OBFD SBSs. This behavior is evident from the increasing co-tier interference at the access and backhaul links of OBFD SBSs. Although a CN cannot support more than $\mathcal{S}$ SBSs,  the co-tier  interference can continue to increase even when $\mathcal{N}_s$ becomes greater than $\mathcal{S}$ since the distance of $\mathcal{S}$ selected SBSs can continue to reduce with the increase in $\lambda^\prime_s$. 
\begin{figure*}
\centering
\begin{minipage}{.5\textwidth}
  \centering
  \includegraphics[width = 3.3in]{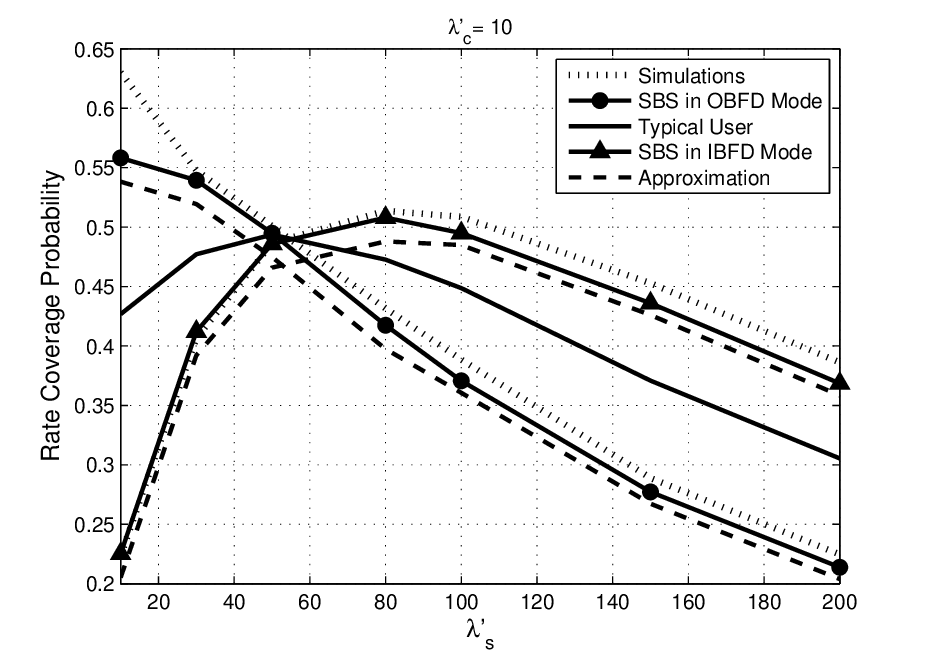}
\caption*{(a)}
\end{minipage}%
\begin{minipage}{.5\textwidth}
  \centering
  \includegraphics[width = 3.3in]{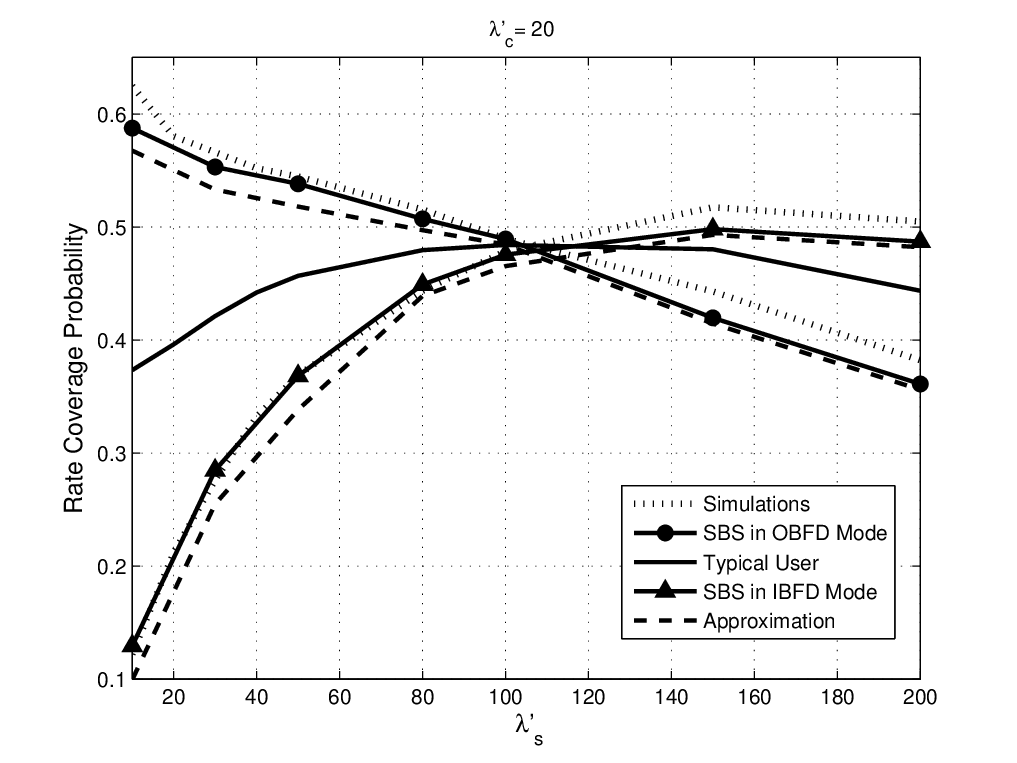}
\caption*{(b)}
\end{minipage}
\caption{Rate coverage probability of SBSs operating in IBFD mode and OBFD mode as well as a typical user in the downlink as a function of the intensity of SBSs considering two different intensities of CNs (for $q=0.5$): (a)~$\lambda_c'=10$ and (b)~$\lambda_c'=20$.}
\label{fig2}
\end{figure*}

In contrast to OBFD SBSs,  the rate coverage of an IBFD SBS is observed to increase first up to some point and then decrease as $\lambda^\prime_s$ increases. That is, for a given proportion $q$ of IBFD SBSs, the number of SBSs in the network  can be optimized to enhance the rate coverage of IBFD SBSs. 
{An increase in the rate coverage is  observed first  due to relatively small backhaul interference incurred at a typical user. However, after a certain point, the backhaul interference  becomes more significant and the rate coverage starts to decrease.}

The trends remain same for different intensities of CNs.
However, it can be observed from \figref{fig2}(b) that a high intensity of CNs is preferable for OBFD mode due to the possibility of strong backhaul links. On the other hand,  an increased intensity of CNs yields a higher backhaul interference in IBFD mode. Consequently, as $\lambda_c'$ increases,  the optimal intensity of SBSs for IBFD mode also increases. It can thus be concluded that a high intensity of SBSs and CNs favors the  IBFD mode and the OBFD mode, respectively.

\subsection{Impact of SI Cancellation}
 \figref{fig3} depicts the impact of increasing the SI cancellation on the performance of SBSs in IBFD mode. As expected, an increase in the SI cancellation  increases the rate coverage of SBSs in IBFD mode whereas the performance of the SBSs in OBFD mode remains unchanged.
For low intensities of SBSs in the network, a higher SI cancellation value may be required to achieve the rate coverage gains of IBFD over OBFD mode. {This is because the weak  access links (due to sparse and thus far-away SBSs) require high transmit power and consequently high SI cancellation is required at the SBSs to achieve high rate coverage.} As such, in scenarios with low $\lambda^\prime_s$, it may not be possible to achieve the gains of IBFD over OBFD mode as is also evident from \figref{fig1} and \figref{fig2}. The OBFD-only mode will thus be the right-choice in such scenarios.

\begin{figure}[h]
\centering
  \includegraphics[width = 3.3in]{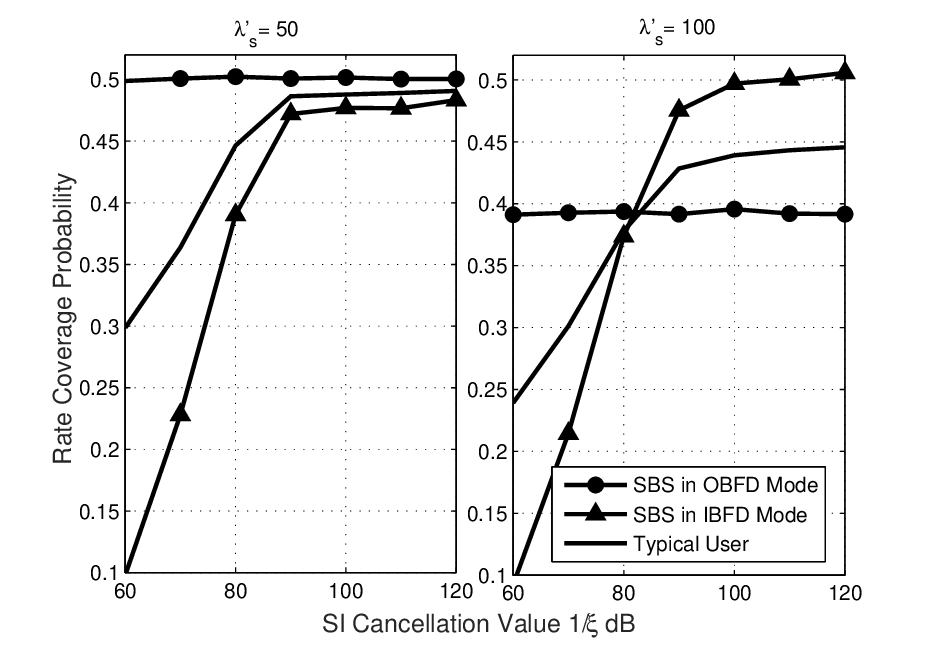}
\caption{Rate coverage probability of SBSs operating in IBFD mode and OBFD mode as well as a typical user in the downlink as a function of the SI cancellation value considering two different intensities of SBSs in the network (for $q=0.5$, $\lambda^\prime_c=10$): (a)~$\lambda^\prime_s=50$ and (b)~$\lambda^\prime_s=100$.}
\label{fig3}
\end{figure}

\begin{figure}[h]
\centering
  \includegraphics[width = 3.4in]{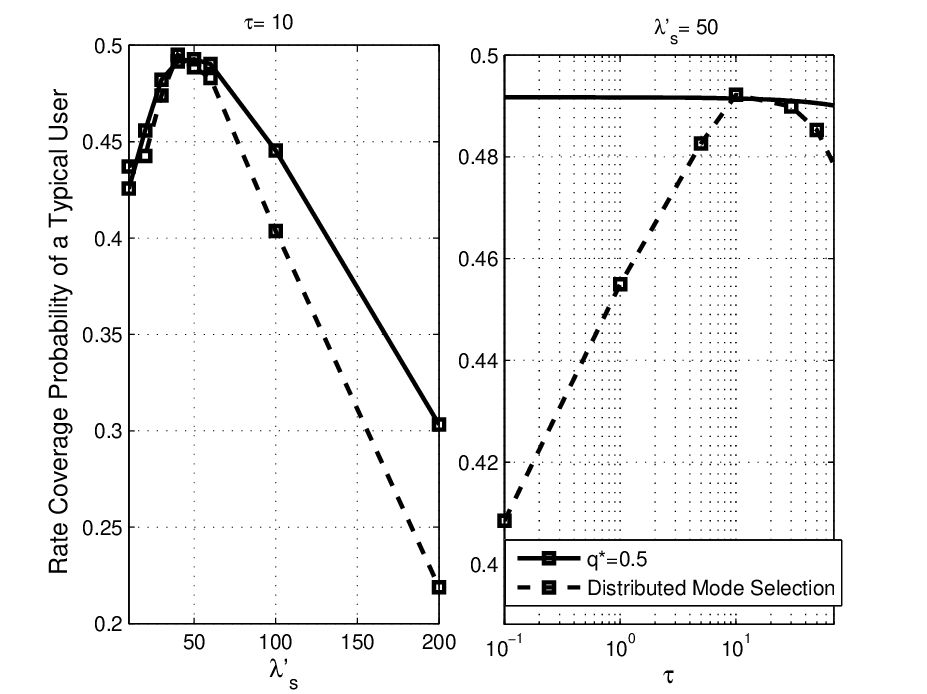}
\caption{Rate coverage probability of a typical user in the downlink with distributed mode selection (for $q^*=0.5$, $\lambda^\prime_c=20$).}
\label{fig5}
\end{figure}

\begin{figure}[h]
\centering
  \includegraphics[width = 3.3in]{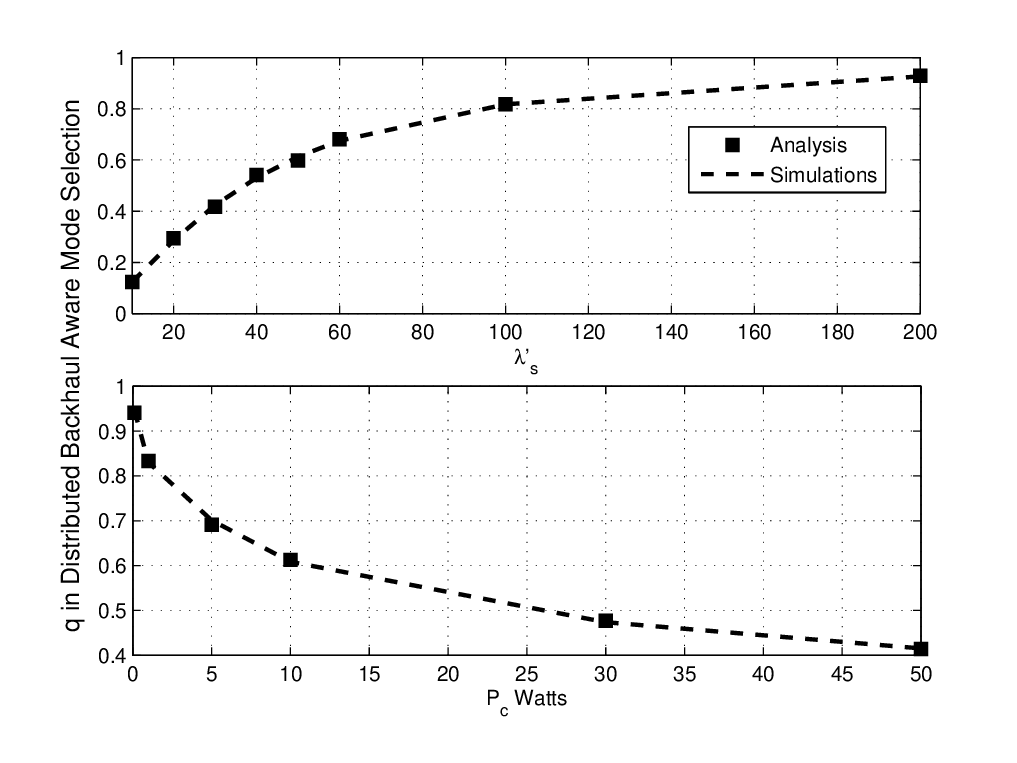}
\caption{Value of $q$ in distributed backhaul-aware mode selection as a function of $P_c$ and $\lambda^{\prime}_s$ ($\lambda^\prime_c=20$).}
\label{rev}
\end{figure}

{\subsection{Distributed Mode Selection}
\figref{fig5} illustrates the impact of distributed mode selection on the rate coverage of a typical user. For comparison, we consider $q^*=0.5$ which is the optimal proportion of IBFD and OBFD SBSs when $\lambda^{\prime}_c=10$ and $\lambda^{\prime}_s=50$ as shown in \figref{fig1}(a). It can be clearly observed that the optimal value of $\tau$ is 10 for $\lambda^{\prime}_s=50$. With this value, the distributed mode selection achieves near optimal performance at $\lambda^{\prime}_s=50$ and other low values of $\lambda^{\prime}_s$. However, the threshold value $\tau=10$ does not work optimally for higher values of $\lambda^{\prime}_s$. The value of $\tau$ will therefore need to be optimized for different network parameters.}
{\figref{rev} depicts the value of $q$ derived in (40) as a function of CN's transmit power and the intensity of SBSs. Simulations results validate the accuracy of the expression. It can be observed that the higher transmit power of CNs lead to higher backhaul interference therefore the proportion of IBFD mode, i.e., $q$ tends to reduce with increase in $P_c$. However, the higher intensities of SBSs favor IBFD mode relatively more compared to OBFD mode as is also depicted in Fig.~3. Therefore, the value of $q$ increases for higher values of $\lambda^{\prime}_s$. These insights can also be depicted directly from (40).}

\subsection{Backhaul Interference Mitigation}
\begin{figure}[h]
\centering
    \includegraphics[width = 3.3in]{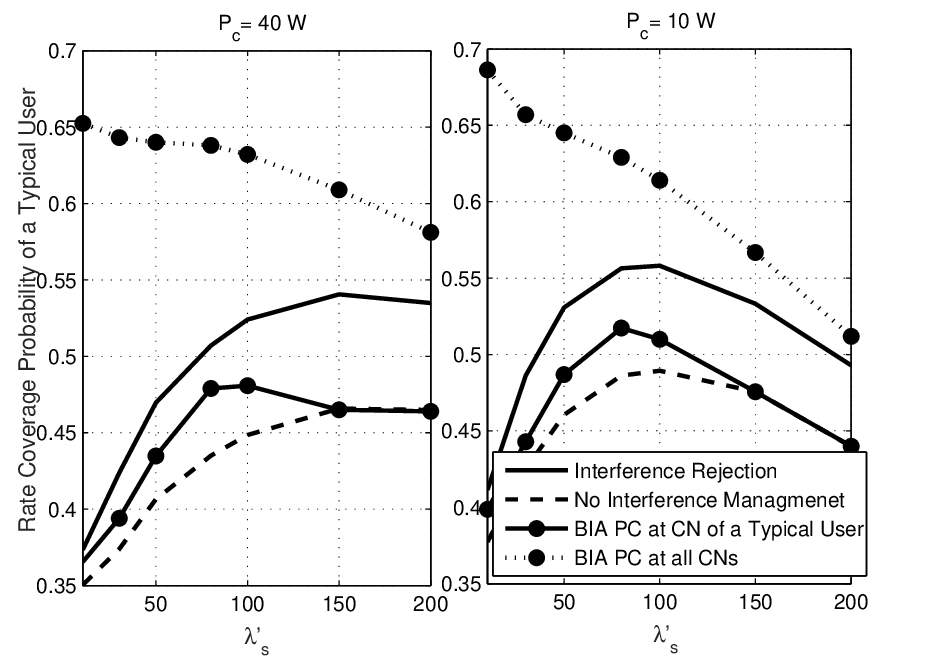}
\caption{Rate coverage probability of  a typical user with interference mitigation schemes implemented at IBFD SBSs  considering two different CN's transmit power (for $q=0.5$, $\lambda_c^{\prime}=20$).}
\label{fig4}
\end{figure}

\begin{figure}[h]
\includegraphics[width = 3.3in]{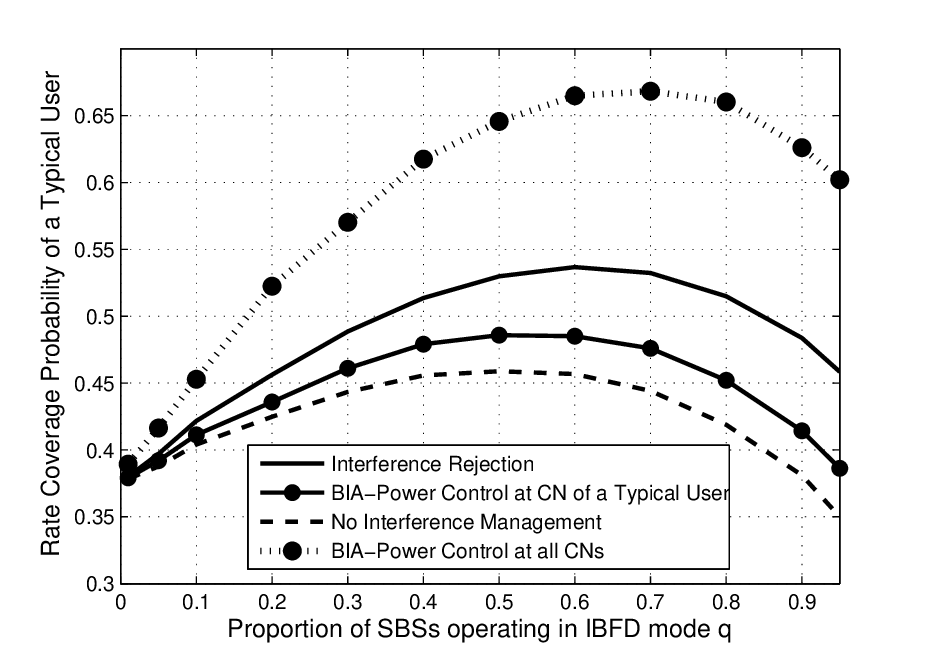}
\caption{Rate coverage probability of  a typical user with interference mitigation schemes implemented at IBFD SBSs  as a function of the proportion of IBFD and OBFD SBSs $q$.}
\label{fig6}
\end{figure}

\figref{fig4} depicts the  rate coverage of a typical user with interference mitigation schemes (i.e., interference rejection and BIA power control) considering different values of the CN's transmit power. As the available transmit power per CN increases, the gains of interference mitigation schemes become more evident. This is due to the fact  that  high  transmit power leads to high backhaul interference, and therefore,  the benefit of interference mitigation becomes more evident.  It can also be observed that if the BIA power control scheme is applied only to the designated CN which is associated with the SBS of the typical user, significant gains can be achieved. In fact, for low intensity of SBSs, the rate coverage becomes similar to that due to the interference rejection scheme. This is because the impact of backhaul interference from neighboring CNs is relatively low compared to the designated CN. However, for high intensity of SBSs, the gains are negligible due to the  significant backhaul interference from neighboring CNs.
Finally, it is observed that once the power control is applied by  all CNs  for their corresponding SBSs, the rate coverage of a typical user enhances significantly. The reason is the reduction in cross-tier interference  from all neighboring CNs.

\figref{fig6} quantifies the gain of implementing interference mitigation schemes at IBFD SBSs as a function of $q$. As expected, the higher the fraction of IBFD SBSs, the gains of interference mitigation  become more evident. Also, the optimal value of $q$ increases with interference mitigation  due to the enhancement in IBFD mode.
 
\subsection{Extensions to Include the Impact of Pilot Contamination}
In order to include the impact of pilot contamination, we need to consider additional interference at the backhaul of all SBSs. The interference due to pilot contamination is received from those CNs that are reusing the same pilot sequences. The interfering CNs follow a thinned PPP of $\Phi_c$ and the intensity $\Phi_c$ can be given as mentioned below.

{Assume that a set of $\mathcal{S}$ orthogonal pilot sequences is reused at each CN.
In case where the number of associated SBSs at a given CN  is less than the number of supported streams, i.e., $\mathcal{N}_s < \mathcal{S}$, the probability of  the use of a pilot sequence at that CN and in turn receiving interference from that CN is $\frac{\mathcal{N}_s}{\mathcal{S}}$. Otherwise, all the pilot sequences will be reused at that CN; thus the probability of receiving interference from that CN is unity.}  The probability of receiving interference from a  CN can thus be given as $\mathbb{P}(\mathcal{N}_s \geq \mathcal{S})+\frac{\mathcal{N}_s}{\mathcal{S}} \mathbb{P}(\mathcal{N}_s < \mathcal{S})$. Consequently, the PPP of interfering CNs  $\bar{\Phi}_{\mathrm{c}}$ with  intensity $\bar{\lambda}_{\mathrm{c}}$ can be obtained by thinning the PPP $\Phi_c$, i.e.,  $\bar{\lambda}_{\mathrm{c}}= \lambda_c \left(\mathbb{P}(\mathcal{N}_s \geq \mathcal{S})+\frac{\mathcal{N}_s}{\mathcal{S}} \mathbb{P}(\mathcal{N}_s < \mathcal{S})\right)$. Since the PMF of $\mathcal{N}_s$ is given by \eqref{pmf}, $\bar{\lambda}_{\mathrm{c}}$ can be given as follows:
\begin{align}
\bar{\lambda}_{\mathrm{c}}&= 
\lambda_c \left(1- \sum_{n=0}^\mathcal{S} \mathbb{P}(\mathcal{N}_s =n)
+ \sum_{n=0}^\mathcal{S} \frac{n}{\mathcal{S}} \mathbb{P}(\mathcal{N}_s =n) \right),
\nonumber \\&
=
\lambda_c \left(1 - \sum_{n=0}^\mathcal{S} \left(1-\frac{n}{\mathcal{S}}\right) \mathbb{P}( \mathcal{N}_s = n)\right).
\end{align}
Consequently, the interference due to pilot contamination at IBFD and OBFD SBSs can be modeled as follows:~\cite[Appendix~A]{bethanabhotla2014optimal}:
\begin{equation}
I_{\mathrm{PC}}=\sum_{y\in \bar{\Phi}_{c}\setminus \mathrm{c} }
\frac{M-\mathrm{min}(\mathcal{N}_s, \mathcal{S})+1}{\mathrm{min}(\mathcal{N}_s, \mathcal{S})}  {P}_c r_{{\mathrm{y}, \mathrm{s}}}^{-\beta}.
\end{equation}
The Laplace transform of $I_{\mathrm{PC}}$ and in turn the rate coverage probability analysis can be conducted using tools similar to those used in  Section~IV.

\subsection{Random Scheduling at a Given SBS}

For a typical random/round-robin scheduling, our framework can be extended by defining the channel access probability of  users attached to a given SBS. The distribution of the number of users per SBS $\mathcal{N}_u$ can be given using  Eq. (3)  by replacing $\mathcal{N}_s$ with $\mathcal{N}_u$ and $\mathbb{E}[\mathcal{N}_u]=\lambda_u/\lambda_s$. Conditioned on $\mathcal{N}_u$, the channel access probability can be calculated as $1/\mathcal{N}_u$ since each user has equal probability to access the channel. The achievable rate of a given user  can then be derived by multiplying the normalized rate in Eq. (7)  with $1/\mathcal{N}_u$ followed by an averaging over $\mathcal{N}_u$.

\section{Conclusion}
We have investigated the performance of a  massive MIMO-enabled wireless backhaul network which is composed of  a mixture of small cells configured either in the in-band or out-of-band FD backhaul mode.
{The feature of massive MIMO at CNs and shared-antenna based full-duplexing at SBSs can enable the use of the proposed framework in existing LTE-A standards.} 
Downlink coverage probability has been derived for a typical user considering both the IBFD and OBFD modes.  It has been shown  that selecting a correct proportion of out-of-band small cells in the network and appropriate  SI cancellation value is crucial in obtaining a high rate coverage.  Few remedial solutions for backhaul interference management  have been presented.  The framework can be extended {to include multiple antennas at SBSs}, to consider the possibility of serving users through CNs, i.e., depending on the coverage requirements a user can opportunistically switch between SBSs and CNs. Further extensions to this work could include the effect of opportunistic scheduling on the rate coverage probability.

\bibliography{IEEEfull,References-1}
\bibliographystyle{IEEEtran}

\end{document}